\documentclass[twocolumn,showpacs]{aastex631}

\usepackage{graphicx}

\usepackage{natbib}
\usepackage{hyperref}
\usepackage{bm}% bold math

\def\sec#1{Sec.~\ref{#1}}

\def\Eq#1{Eq.~(\ref{#1})}

\def\eqref#1{(\ref{#1})}

\def\sec#1{Sec.~\ref{#1}}

\newcommand{\e}{\ensuremath{\textrm{e}}}
\newcommand{\td}{\ensuremath{\textrm{d}}}

\begin{document}
\title{Identifying the QCD Phase Transitions via the Gravitational Wave Frequency \\ from Supernova Explosion}

\author{Zhan Bai}
\email{baizhan@pku.edu.cn}
\affiliation{Department of Physics and State Key Laboratory of Nuclear Physics and Technology, Peking University, Beijing 100871, China}

\author{Wei-jie Fu}
\email{wjfu@dlut.edu.cn}
\affiliation{Institute of Theoretical Physics, School of Physics \&
   Optoelectronic Technology, \\
	  Dalian University of Technology, Dalian, 116024,
  P.R. China}

\author{Yu-xin Liu}
\email{yxliu@pku.edu.cn}
\affiliation{Department of Physics and State Key Laboratory of Nuclear Physics and Technology, Peking University, Beijing 100871, China}
\affiliation{Collaborative Innovation Center of Quantum Matter, Beijing 100871, China}
\affiliation{Center for High Energy Physics, Peking University, Beijing 100871, China}

\begin{abstract}
We investigate the non-radial oscillations of newly born neutron stars (NSs) and strange quark stars (SQSs).
This is done with the relativistic nuclear field theory with hyperon degrees of freedom employed to describe the equation of state for the stellar matter in NSs,
and with both the MIT bag model and the Nambu--Jona-Lasinio model adopted to construct the configurations of the SQSs.
We find that the gravitational-mode ($g$-mode) eigenfrequencies of newly born SQSs are significantly lower than those of NSs,
which is independent of models implemented to describe the equation of state for the strange quark matter.
Meanwhile, the eigenfrequencies of the other modes of non-radial oscillations, {\it e.g.}, fundamental ($f$)- and pressure ($p$)-modes, are much larger than those of the $g$-mode, and is related to the stiffness of the equation of states (EoSs).
In the light of the first direct observation of gravitational waves,
it is promising to employ the gravitational waves to identify the QCD phase transition in high density strong interaction matter.
\end{abstract}

%\pacs{25.75.Nq, %Quark deconfinement, quark-gluon plasma production, and phase transitions
%      04.30.Tv, %Gravitational-wave astrophysics
%      26.60.Kp, %Equations of state of neutron-star matter
%      97.60.Bw  %Supernovae
%     }                             % Classification Scheme.

%\maketitle

\section{\label{sec:Intro}Introduction}
The first direct observation of gravitational waves (GW) from a binary black hole (BH) merger by LIGO Scientific Collaboration and Virgo Collaboration~\citep{LIGOScientific:2016aoc},
has opened up a new era for the astronomy, viz. the gravitational wave astronomy.
In 2017, another gravitational wave from binary neutron star coalescence was first detected~\citep{LIGOScientific:2017vwq}.
This gravitational wave was also associated with a short gamma-ray burst~\citep{LIGOScientific:2017zic},
and marks the dawn of multi-messenger astronomy.
Up to now, all the detected GW events are compact binary mergers~\citep{LIGOScientific:2018mvr,LIGOScientific:2020ibl,LIGOScientific:2021tsm}.

However,
 the potential high-frequency sources sought by earth-based gravitational-wave detectors, such as LIGO, VIRGO, GEO600, TAMA300,
include not only the binary merger and ringdown,
but also the tidal disruption of a neutron star (NS) by its companion BH in NS/BH binaries, spinning NSs,
type-II supernovae, proto-neutron-stars produced by the accretion-induced collapse of white dwarf stars, etc.~\citep{Cutler:2002me,Andersson:2010ufc,Haskell:2021ljd},
One of the most prominent source of GW is the core-collapse supernova (CCSN)~\citep{Gossan:2015xda,Szczepanczyk:2021bka}.

On the other hand, NS, as one of the most compact objects in the cosmos, is an ideal laboratory for physics in extreme environments,
such as the strong gravitational fields, QCD at high density, high magnetic fields, etc., see, {\it e.g.}, \citet{Glendenning:2000} for details.
In fact, there have been a longstanding debate whether the QCD phase transition takes place in the interior of these compact stars,
{\it i.e.}, whether strange quark stars (SQSs) are also potential candidates for them
~\citep{Itoh:1970uw,Freedman:1977gz,Witten:1984rs,Alcock:1986hz,Weber:2004kj,Alford:2006vz,Lattimer:2006xb}.
To answer this question, one can, of course, resort to earth-based heavy-ion collision experiments,
such as at the Relativistic Heavy-Ion Collider (RHIC)~\citep{STAR:2005gfr,PHENIX:2004vcz}, the Large Hadron Collider (LHC) at CERN~\citep{ALICE:2010suc},
the Facility for Antiproton and Ion Research (FAIR) at GSI and the Nuclotron-based Ion Collider fAcility (NICA) at Dubna~\citep{Batyuk:2016yjp}.
Indeed, significant progresses have been made in recent years from experimental measurements, for instance,
the Beam Energy Scan (BES) program at RHIC ~\citep{STAR:2013gus,STAR:2014egu,Luo:2015ewa},
aiming to get hold of the existence and location of the critical end point (CEP) in the QCD phase diagram~\citep{Luo:2015ewa,Stephanov:2006zvm}.
The QCD phase structure, however, is far from being unveiled, because of the notorious sign problem at finite chemical potential,
which prevents the first-principle lattice QCD simulations from getting access to high density regime~\citep{Aarts:2015tyj}.
In turn, the gravitational wave astronomy provides a potential, new approach to study the QCD at extreme densities.

Many observable differences have been proposed to identify the QCD phase transition in compact stars, {\it i.e.}, to distinguish SQSs from NSs.
One of the most important observable is mass of the compact star.
Several compact stars with mass larger than $2M_{\odot}$ have already been observed~\citep{Demorest:2010bx,Antoniadis:2013pzd,Fonseca:2016tux,NANOGrav:2017wvv,Linares:2018ppq,NANOGrav:2019jur},
which requires that the equation of state (EoS) to be stiff, and provides constraints on dense matter models.
Also, SQSs are found to have larger dissipation rate of radial vibrations~\citep{Wang:1984edr} and higher bulk viscosity~\citep{Haensel:1989AA}.
The cooling of SQSs is more rapid than that of NSs within the first 30 years~\citep{Schaab:1997hx}.
It was found that the spin rate of SQSs can be much closer to the Kepler limit than that of NSs~\citep{Madsen:1992sx}.
Recently, it has been found that the observation of old SQSs can set important limits on the scattering cross-sections between the light quarks and the bosonic non-interacting dark matter
~\citep{Zheng:2016ygg}.

As for the signal from GW,
the detection of gravitational wave from binary NS coalescence can provide constraints on the EoS,
and can serve as an indicator of the hadron-quark phase transition
~\citep{Bauswein:2009im,Annala:2017llu,Fattoyev:2017jql,Nandi:2017rhy,De:2018uhw,Bauswein:2018bma}.
There has also been plenty of studies of GW from single mature neutron star,
which can be used to detect the appearance of super-fluid~\citep{Passamonti:2015oia,Kantor:2014lja,Yu:2016ltf,Yu:2017cxe},
the effect of hyperon~\citep{Dommes:2015wul},
the possible color-flavor locked phase~\citep{Flores:2017hpb},
and hadron-quark phase transition~\citep{Sotani:2010mx,Wei:2018tts,Flores:2013yqa,Jaikumar:2021jbw,Ranea-Sandoval:2018bgu}.
However, as for the GW from proto-neutron star (PNS) after supernova explosion, the relevant study is still in yearning.

In our previous work~\citep{Fu:2008bu}, we have studied the eigenfrequencies of the gravitational mode,
{\it i.e.}, the $g$-mode, oscillation of newly born SQSs and NSs,
and found that the eigenfrequencies of SQSs are much lower than those of NSs,
because the components of a SQS are almost extremely relativistic particles while nucleons in a NS are non-relativistic.
Furthermore, simulations of core-collapse supernovae have indicated that $g$-mode oscillations of the supernovae core may be excited~\citep{Burrows:2005dv,Burrows:2006ci},
and serve as efficient sources of gravitational waves~\citep{Ott:2006qp,Owen:2005fn,Lai:2006pr},
for more related discussions about the simulations of core-collapse supernovae, see {\it e.g.}~\citet{Blondin:2002sm,Foglizzo:2001ke}.

Apart from the $g$-mode, oscillation from other modes, such as $f$- and $p$-mode,
are important in the study of compact stars.
For example, the $f$-mode frequency is related to some universal relations
(compact star quantities that are irrelevant to the EoS)~\citep{Chan:2014kua,Chirenti:2015dda},
and the $f$-mode GW is a powerful tool for inferring the compact stars mass, radius and moment of inertia~\citep{Lau:2009bu},
and can be used to study the color-flavor locked state inside compact stars~\citep{Flores:2017kte}.
The different oscillation modes can be coupled with the tide during coalescence,
and the phase of GW may be altered~\citep{Zhou:2018tvc}.
Therefore, in this work we will extend our studies in ~\citet{Fu:2008bu} to study the $f$- and $p$-mode oscillation for comparison.
For NS, we will also consider the hyperon degrees of freedom for EoS of the matter.

In ~\citet{Fu:2008bu}, the quark matter was studied in the framework of MIT bag model,
where all the quarks are treated as free fermions with small masses.
However, the quarks might acquire a large constituent mass through strong interaction, known as dynamical chiral symmetry breaking (DCSB),
and $g$-mode oscillation frequency might be altered by this constituent mass.
Therefore, in this paper, we study the oscillation in the framework of Nambu--Jona-Lasinio model,
	which takes into consideration the DCSB effect.

The observed large mass also requires the EoS for dense matter to be stiff.
In this paper, we consider both the stiff hadron model and  the stiff quark model,
and see whether they will change oscillation frequencies significantly.

Since our main interest is to detect the GW from NS and SQS oscillation,
it is necessary to study the damping of the oscillation through GW emission.
If the damping time is small, then most of the oscillation energy will be transformed into GW,
and it might be possible to observe on earth.
On the contrary, if the damping time is large,
most of the energy will be transformed into other channels and the GW will hardly be detected.
Therefore, in this work, we will calculate the damping time through GW for different models.

Details about the calculations will be presented, and the dependence of the eigenfrequencies on different equations of state will also be discussed.
The paper is organized as follows. In Sec.~\ref{sec:nonradial} the nonradial oscillations of a nonrotating, unmagnetized, and fluid star are briefly described.
In \sec{sec:RMF:MIT}, the model of newly born NSs and SQSs is constructed using RMF model and MIT bag model, respectively,
and some numerical results and discussions are presented.
In \sec{sec:NJL} the NJL model is adopted to describe the EoS of the strange quark matter,
and relevant calculated results of the $g$-mode eigenfrequencies are presented.
In Sec.~\ref{sec:stiffEOS} we present the result obtained with both stiff RMF model and stiff NJL model.
In Sec.~\ref{sec:damping}, the damping times through GW emission for different models are calculated.
Section~\ref{sec:sum} summarizes our conclusions.

\section{\label{sec:nonradial}Nonradial Oscillations of a Star}
%\subsection{Oscillation}
In this section, we reiterate briefly the nonradial oscillations of a nonrotating, unmagnetized, and fluid star (for details, see {\it e.g.}, \citet{Reisenegger:1992APJ,Cox:1980:Theory_of_Stellar_Pulsations,Lai:1993di}).
Assuming the equilibrium configuration of the star to be spherically symmetric,
the pressure $p_{0}^{}$, density $\rho_{0}^{}$ and the gravitational potential $\phi_{0}^{}$ are functions of only the radial coordinate $r$.
A vector field of displacement $\bm{\xi}(\bm{r},t)$ describes the oscillations, and the Eulerian perturbations of the pressure, density and the potential are $\delta p$, $\delta \rho$ and $\delta \phi$, respectively.
In order to simplify the calculation, we use Newtonian equations in this paper.
The correction due to general relativity is expected to be in the order of $GM/(Rc^2)\sim 10$ per cent.
For single cold mature neutron star,
there have been calculations showing that Newtonian and relativistic results are in qualitatively agreement,
see \citet{Passamonti:2008sb} and \citet{Gaertig:2009rr},
or \citet{Yu:2016ltf} and \citet{Yu:2017cxe} for comparison.
The effect of general relativity will be discussed in our future work.

The Newtonian equation of motion (EoM) reads
\begin{equation}
-\nabla p-\rho\nabla\phi=\rho\frac{\partial^{2}\bm{\xi}}{\partial t^{2}} \, ,\label{eq:eom1}
\end{equation}
where $p=p_{0}^{}+\delta p$, as well as $\rho$ and $\phi$. Assuming the perturbations are small, and only keeping linear terms of the perturbations, we are led to
\begin{equation}
-\nabla \delta p - \delta \rho\nabla\phi_{0}^{} - \rho_{0}^{}\nabla \delta
\phi = \rho_{0}^{} \frac{\partial^{2}\bm{\xi}}{\partial t^{2}} \, ,      \label{eq:eom2}
\end{equation}
where we have employed equilibrium equation, {\it i.e.},
\begin{equation}
-\nabla p_{0}-\rho_{0}\nabla\phi_{0}=0 \, .   \label{eq:equi}
\end{equation}
In a slight abuse of notation we employ $p$, $\rho$ and $\phi$ to denote the equilibrium values from now on, rather than those with subscript $_0$.

Furthermore, we have continuity equation which reads
\begin{equation}
\delta \rho+\nabla\cdot(\rho\bm{\xi}) = 0 \, . \label{eq:continuity}
\end{equation}
Here, it is more convenient to adopt the Lagrangian perturbations $\Delta$, which is related with the Eulerian formalism through the following relation, {\it i.e.},
\begin{equation}
\Delta=\delta+\bm{\xi}\cdot\nabla\,.
\end{equation}
Here, it is more convenient to adopt the Lagrangian perturbations $\Delta$, which is related with the Eulerian formalism through the following relation, {\it i.e.},
\begin{equation}
\Delta=\delta+\bm{\xi}\cdot\nabla\,.
\end{equation}
Then Eq.~(\ref{eq:continuity}) follows readily as
\begin{equation}\label{eq:continuity2}
\Delta \rho+\rho\nabla\cdot\bm{\xi}=0 \, .
\end{equation}

For a stellar oscillation with an eigenfrequency $\omega$, all the perturbative quantities, {\it i.e.}, $\bm{\xi}$, $\delta p$, $\delta\rho$, and $\delta \phi$,
are dependent on time through a temporal factor $e^{-i\omega t}$.
Inserting this factor into the EoM (in Eq.~(\ref{eq:eom2})), we arrive at
\begin{equation}
-\omega^{2}\bm{\xi}=-\frac{\nabla(\delta p)}{\rho}-\frac{\delta
\rho}{\rho}\nabla\phi-\nabla(\delta\phi)\, .\label{eq:eom3}
\end{equation}
Furthermore, we have the linearized Poisson equation for the gravitational potential, {\it i.e.},
\begin{equation}
\nabla^{2}\delta\phi=4\pi G \delta\rho \, ,  \label{eq:Poisson}
\end{equation}
where $G$ is the gravitational constant. Eqs.~(\ref{eq:continuity2})--(\ref{eq:Poisson}) constitute a closed set of equations of nonradial oscillations for a nonrotating, fluid star.
In the following we will rewrite those vector equations in component formalism.
The transversal component of displacement ${\bm{\xi}}_{\bot}$, {\it i.e.}, perpendicular to the radial direction, is easily deduced from Eq.~(\ref{eq:eom3}), which reads
\begin{equation}
{\bm{\xi}}_{\bot}=\frac{\nabla_{\bot}(\delta
p)}{\omega^{2}\rho}+\frac{\nabla_{\bot}(\delta\phi)}{\omega^{2}}\,,\label{eq:xiPer}
\end{equation}
with
\begin{equation}
\nabla_{\bot}=\frac{1}{r}\frac{\partial}{\partial\theta}\bm{e}_{\theta}
+\frac{1}{r\sin\theta}\frac{\partial}{\partial\varphi}\bm{e}_{\varphi}\,.
\end{equation}
Substituting Eq.~(\ref{eq:xiPer}) into Eq.~(\ref{eq:continuity2}), one arrives at
\begin{equation}
\frac{\Delta \rho}{\rho}+\frac{1}{r^{2}}\frac{\partial}{\partial
r}(r^{2}\xi_{r})+\frac{1}{\omega^{2}\rho}\nabla_{\bot}^{2}(\delta
p)+\frac{1}{\omega^{2}}\nabla_{\bot}^{2}(\delta \phi)=0  \,  ,\label{eq:continuity3}
\end{equation}
with
\begin{equation}
\nabla_{\bot}^{2}=\frac{1}{r^{2}\sin\theta}\frac{\partial}{\partial\theta}\left(\sin\theta\frac{\partial}{\partial\theta}\right)
+\frac{1}{r^{2}\sin^{2}\theta}\frac{\partial^{2}}{\partial\varphi^{2}}\, .
\end{equation}

Both the $\delta p$ and $\delta\phi$  in an eigenmode of the oscillations can be factorized into products of a spherical harmonic $Y_{lm}(\theta,\varphi)$  and a radial coordinate dependent function.
Thus, for an eigenmode, Eq.~(\ref{eq:continuity3}) is simplified to the following equation:
\begin{equation}
\frac{1}{r^{2}}\frac{\partial}{\partial
r}(r^{2}\xi_{r})-\frac{l(l+1)}{\omega^{2}r^{2}}\frac{\delta
p}{\rho}-\frac{l(l+1)}{\omega^{2}r^{2}}\delta \phi+\frac{\Delta
\rho}{\rho}=0\, .\label{eq:continuity4}
\end{equation}

Considering the adiabatic sound speed $c_{s}$ defined as
\begin{equation}
c_{s}^{2}\equiv\left(\frac{\partial p}{\partial
\rho}\right)_{\mathrm{adia}}=\frac{\Delta p}{\Delta \rho} \, , \label{eq:cs}
\end{equation}
one has
\begin{eqnarray}
\Delta \rho & = & \frac{\Delta p}{c_{s}^{2}}=\frac{(\delta+\bm{\xi}\cdot\nabla)p}{c_{s}^{2}}\nonumber\\
& = & \frac{\delta p-\xi_{r}g\rho}{c_{s}^{2}}\,,\label{eq:Delrho}
\end{eqnarray}
where we have employed the local gravitational acceleration $\bm{g}\equiv-\nabla\phi$.
Inserting Eq.~(\ref{eq:Delrho}) into Eq.~(\ref{eq:continuity4}), we are left with
\begin{eqnarray}
\frac{\partial}{\partial r}(r^{2}\xi_{r}) =\, \frac{g}{c_{s}^{2}}(r^{2}\xi_{r})&+&\left[\frac{l(l+1)}{\omega^{2}}-\frac{r^{2}}{c_{s}^{2}}\right]\left(\frac{\delta p}{\rho}\right)\nonumber\\
&+&\frac{l(l+1)}{\omega^{2}}\delta\phi\,\label{eq:Oscil1}.
\end{eqnarray}

The radial component of the EoM in Eq.~(\ref{eq:eom3}) is readily obtained, which is given by
\begin{equation}
-\omega^{2}\xi_{r}=-\frac{1}{\rho}\frac{\partial(\delta p)}{\partial
r}-\frac{\delta \rho}{\rho}g-\frac{\partial(\delta\phi)}{\partial
r} \, .\label{eq:radial}
\end{equation}

For the $\delta \rho$, one has
\begin{eqnarray}
\delta \rho& = & (\Delta-\bm{\xi}\cdot\nabla)\rho=\frac{\Delta p}{c_{s}^{2}}-\xi_{r}\frac{d \rho}{d r}\nonumber\\
& = &\frac{\Delta p}{c_{s}^{2}}-\frac{\xi_{r}}{c_{e}^{2}}\frac{d p}{d r}\nonumber\\
& = &\left(\frac{\rho}{c_{s}^{2}}\right)\left(\frac{\delta p}{\rho}\right)+g^{2}\left(\frac{1}{c_{e}^{2}}-\frac{1}{c_{s}^{2}}\right)\left(\frac{\rho}{g r^{2}}\right)(r^{2}\xi_{r})  ,\,  \label{eq:drho}
\end{eqnarray}
where we have introduced another sound speed, {\it i.e.}, equilibrium sound speed $c_{e}$, defined by
\begin{eqnarray}
c_{e}^{2}\equiv\left(\frac{d p}{d
\rho}\right)_{\mathrm{equi}}=\frac{d p/d r}{d \rho/d r}  \,  .   \label{eq:ce}
\end{eqnarray}
In the same time, we have
\begin{eqnarray}
\frac{\partial(\delta p)}{\partial r}
= \rho\frac{\partial}{\partial r}\left(\frac{\delta p}{\rho}\right) -
\frac{g\rho}{c_{e}^{2}}\left(\frac{\delta p}{\rho}\right)    \,    .   \label{eq:ddelp}
\end{eqnarray}
Substituting Eqs.~(\ref{eq:drho}) and (\ref{eq:ddelp}) into Eq.~(\ref{eq:radial}), one is led to
\begin{eqnarray}
\frac{\partial}{\partial r}\left(\frac{\delta p}{\rho}\right) =&&\frac{\omega^{2}-\omega_{BV}^{2}}{r^{2}}(r^{2}\xi_{r})\nonumber\\
&+&\frac{\omega_{BV}^{2}}{g}\left(\frac{\delta p}{\rho}\right)-\frac{\partial \delta\phi}{\partial r}   ,\,
\label{eq:Oscil2}
\end{eqnarray}
where $\omega_{BV}^{}$ is the Brunt-V\"{a}is\"{a}l\"{a} frequency, which is given by
\begin{eqnarray}
\omega_{BV}^{2}=g^{2}\left(\frac{1}{c_{e}^{2}}-\frac{1}{c_{s}^{2}}\right)
\, .\label{eq:BV}
\end{eqnarray}

In the same way, the Poisson equation for perturbations of the gravitational potential in Eq.~(\ref{eq:Poisson}) can be reformulated as
\begin{eqnarray}
\frac{1}{r^{2}}\frac{\partial}{\partial r}\left(r^{2}\frac{\partial\delta\phi}{\partial r}\right) = && \, 4\pi G\left[\omega_{BV}^{2}\left(\frac{\rho}{gr^{2}}\right)(r^{2}\xi_{r}) \right.\nonumber\\
&&+ \left. \left(\frac{\rho}{c_{s}^{2}}\right)\left(\frac{\delta p}{\rho}\right)\right]+\frac{l(l+1)\delta\phi}{r^{2}} \,
. \qquad \label{eq:Poisson2}
\end{eqnarray}

In the following calculations, we adopt the so-called Cowling approximation~\citep{Cowling:1941MNRAS}, {\it i.e.}, neglecting the perturbations of the gravitational potential $\delta\phi$.
For a non-rotating compact star, the Cowling approximation will only lead to a small difference in the oscillation frequency~\citep{Yoshida:1997bf,Gregorian:2014MASTER}.
The Cowling approximation has about $10\%\sim 30\%$ effect on the $f$-mode oscillation ~\citep{Chirenti:2015dda},
less than $20\%$ effect on the $p$-mode oscillation frequency~\citep{Yoshida:1997bf},
and less than $5\%$ for $g$-mode frequency~\citep{Sotani:2001bb,Xu:2017hqo}.
Therefore, we assume that the Cowling approximation is tolerable,
and leave the full relativistic calculation in our future work.

For the convenience of numerical computations in the following, we parameterize the displacement vector as:
\begin{eqnarray}
\bm{\xi}=(\eta_{r}\bm{e}_{r}+r\eta_{\bot}\nabla_{\bot})Y_{lm}(\theta,\varphi)e^{-i\omega
t}   \,  .  \label{eq:eta}
\end{eqnarray}
Recalling Eq.~(\ref{eq:xiPer}), we have
\begin{eqnarray}
\xi_{r}(r,\theta,\varphi,t) & = \eta_{r}(r)Y_{lm}(\theta,\varphi)e^{-i\omega
t},\nonumber\\
\frac{\delta p(r,\theta,\varphi,t)}{r\omega^{2}\rho}& =
\eta_{\bot}(r)Y_{lm}(\theta,\varphi)e^{-i\omega t}   \,  .   \label{eq:displa}
\end{eqnarray}
Inserting these two equations into Eqs.~(\ref{eq:Oscil1}) and (\ref{eq:Oscil2}), we obtain
\begin{eqnarray}
\frac{d \eta_{r}}{d r}\! & = &\left(\frac{g r}{c_{s}^{2}}-2\right)\frac{\eta_{r}^{}}{r}+\left[l(l+1)-\frac{\omega^{2}r^{2}}{c_{s}^{2}}\right]\frac{\eta_{\bot}^{}}{r}  \,  , \label{eq:Osc1} \\[1mm]
\frac{d \eta_{\bot}}{d r}\! & =&\left(1-\frac{\omega_{BV}^{2}}{\omega^{2}}\right)\frac{\eta_{r}^{}}{r}+\left(\frac{\omega_{BV}^{2}r}{g}-1\right)\frac{\eta_{\bot}^{}}{r} \, .
\label{eq:Osc2}
\end{eqnarray}
In order to determine the eigenfrequency of a mode of oscillations, one also needs the boundary conditions which, in the stellar center, is given by
\begin{eqnarray}
\eta_{r}^{} = A\,lr^{l-1}, \qquad \eta_{\bot}^{} = A\,r^{l-1} \qquad (r\rightarrow
0)\,,\label{eq:bound1}
\end{eqnarray}
where $A$ is an arbitrary constant, and is related with the amplitude of the oscillation. While at the outer surface of the star, the Lagrangian perturbation of the pressure is vanishing, {\it i.e.},
\begin{eqnarray}
\Delta p & =&(\delta +\bm{\xi}\cdot\nabla)p\nonumber\\
&=&(\omega^{2}r\eta_{\bot}^{} - g\eta_{r}^{})\rho=0  \qquad (r=R)\,\label{eq:bound2}.
\end{eqnarray}

In our calculations we employ the Newtonian hydrostatic equations, rather than those of general relativity, to determine the equilibrium configurations of newly born NSs and SQSs,
in order to coincide with Eqs.~(\ref{eq:Oscil1}) and (\ref{eq:Oscil2}) or Eqs.~(\ref{eq:Osc1}) and (\ref{eq:Osc2}).
Then, the adiabatic, equilibrium sound speed, $c_{s}$ and $c_{e}$ respectively, and the Brunt-V\"{a}is\"{a}l\"{a} frequency $\omega_{BV}^{}$ can be obtained as functions of the radial coordinate $r$.

Solving the oscillation equations in Eqs.~(\ref{eq:Osc1}) and (\ref{eq:Osc2}),
together with the boundary conditions in Eqs.~(\ref{eq:bound1}) and (\ref{eq:bound2}),
one obtains three classes of modes of oscillations, which are the gravitational-mode ($g$-mode), fundamental-mode ($f$-mode), and the pressure-mode ($p$-mode), respectively.

Generally speaking, stellar oscillations of $g$-mode originate from the buoyancy in the star and,
thus, the eigenfrequency of $g$-mode is intimately linked to Brunt-V\"{a}is\"{a}l\"{a} frequency $\omega_{BV}^{}$.
$f$-mode and $p$-mode, however, result from the pressure inside the star,
and $f$-mode is in fact a particular mode of $p$-modes, with the number of the radial node being zero.
The eigenfrequencies of $f$-mode and $p$-mode are therefore related only with the sound speed of the stellar matter.
Of the three classes of oscillation modes, $g$-mode is our main focus, because it may be a source of the gravitational wave
~\citep{Burrows:2005dv,Burrows:2006ci,Ott:2006qp,Owen:2005fn,Lai:2006pr}.

\section{Modelling the Newly Born Compact Stars}\label{sec:RMF:MIT}

\subsection{Newly Born Neutron Stars --- in RMF model}\label{sec:NSs}

In this section as well as those below, we will construct models of newly born NSs and SQSs, respectively.
In comparison with ordinary NSs after cooling, such as pulsars, newly born NSs in a core-collapse supernova feature two significant properties: one is the high temperature.
The inner temperature of pulsars is of order of $10^9\,\mathrm{K}$ and below, while that of newly born NSs can be higher than $10^{11}\,\mathrm{K}$.
The other is the high abundance of neutrinos or leptons.
In the first tens of seconds after the core bounce in a core-collapse supernova, neutrinos and antineutrinos of different flavors are trapped in the stellar matter of high temperatures and densities,
since their mean free paths are smaller than the size of NSs~\citep{Bethe:1990mw}.
Here we will not delve into the evolution and structure of the newly born NSs, for more details, see, {\it e.g.}, \citet{Burrows:1986me,Pons:1998mm,Keil:1995hw,Prakash:1996xs,Pons:2000xf}.
In this work we extend our former calculations in \citet{Fu:2008bu} to allow for the existence of hyperons,
besides nucleons. A fluid element inside the star can be described by three independent variables,
such as the baryon density $\rho_{B}^{}$, the entropy per baryon $S$, and the lepton fraction $Y_{L}=Y_{e}+Y_{\nu_{e}}$ ($Y_{i}={\rho_{i}^{}}/{\rho_{B}^{}}$).
Surely, one can use the electron fraction $Y_{e}$ in lieu of $Y_{L}$, which are related with each other through beta equilibrium conditions.
In our calculations we employ the relativistic mean field (RMF) theory at finite temperature to describe the interactions among baryons~\citep{Serot:1984ey,Glendenning:2000},
while assume all other components of the star to be non-interacting. The Lagrangian density of the RMF is given by
\begin{eqnarray}
\mathcal{L}  &=  &\sum_i\bar{\Psi}_{i}\left[i\gamma_{\mu}\partial^{\mu}- m_{i}^{\ast} - g_{\omega i} \gamma_{\mu} \omega^{\mu} - {g_{\rho i}^{}} \gamma^{\mu}\bm{t}_i\cdot {\bm{\rho}_{\mu}^{}} \right]\Psi_{i}\nonumber\\
& & + \frac{1}{2}\!\!\left(\partial_{\mu}\sigma\partial^{\mu}\sigma \! - \! m_{\sigma}^{2}\sigma^{2}\right) \! - \! \frac{1}{3}b\,m_{N}^{}({g_{\sigma N}^{}} \sigma)^{3} 
\! - \! \frac{1}{4}c({g_{\sigma N}^{}} \sigma)^{4}\nonumber\\
& & - \frac{1}{4}\omega_{\mu\nu}\omega^{\mu\nu}+\frac{1}{2}m^{2}_{\omega} \omega_{\mu}\omega^{\mu}   \nonumber\\
& & - \frac{1}{4} {\bm{\rho}_{\mu\nu}^{}} \cdot\bm{\rho}^{\mu\nu} +\frac{1}{2}m^{2}_{\rho} {\bm{\rho}_{\mu}^{}} \cdot \bm{\rho}^{\mu},\label{eq:RMFlagrangian}
\end{eqnarray}
where $m_{i}^{\ast}=m_{i}-g_{\sigma i}\sigma$ is the effective mass of the baryon,
$\Psi_{i}$ ($i=p,n,\Lambda,\Sigma^{\pm,0},\Xi^{-,0}$) are the baryon octet fields,
which interact through exchanging mesons with coupling strength $g_{ji}^{}$ ($j$ refers to the meson $\sigma$, $\rho$ and $\omega$, $i$ stands for the baryons).
The isoscalar-scalar meson $\sigma$ provides attractive interactions between baryons, while the isoscalar-vector meson $\omega$ accounts for repulsive interactions.
The isovector-vector $\rho$ meson distinguishes between baryons with different isospin, and thus plays a significant role in determining, {\it e.g.}, the symmetry energy of the nuclear matter.
Note that self-interactions of $\sigma$-meson in Eq.~(\ref{eq:RMFlagrangian}) are also imperative to the correct description of the nuclear matter in the RMF formalism.
$\omega_{\mu\nu}^{}$ and $\bm{\rho}_{\mu\nu}^{}$ in Eq.~(\ref{eq:RMFlagrangian}) are the vector meson field tensors, which read
\begin{eqnarray}
 \omega_{\mu\nu} & \equiv
\partial_{\mu}\omega_{\nu}-\partial_{\nu}\omega_{\mu},\\
 \bm{\rho}_{\mu\nu} & \equiv
\partial_{\mu}\bm{\rho}_{\nu}-\partial_{\nu}\bm{\rho}_{\mu},
\end{eqnarray}
and $t^a$ is the Pauli matrices with $\mathrm{tr}(t^at^b)=\delta_{ab}/2$.
In the mean field approximation, the Dirac equation for baryon $i$ follows readily as
\begin{eqnarray}
\left[i\gamma_{\mu}\partial^{\mu} - m_{i}^{\ast}- {g_{\omega i}^{}} \gamma^{0} {\omega_{0}^{}}   - {g_{\rho i}^{}} \gamma^{0} t_{3i}^{} {\rho_{03}^{}} \right]\Psi_{i}=0,   \label{eq:Dirac}
\end{eqnarray}
and thus, the dispersion relation for baryons reads
\begin{eqnarray}
E_{i}&=\sqrt{p^{2}+{m^{*}_{i}}^{2}} + {g_{\omega i}^{}} \omega_{0}+{g_{\rho i}^{}} {\rho_{03}^{}}t_{3i} \,,
\label{eq:dispersion}
\end{eqnarray}
with  $t_{3i}^{}$  the 3-component of the isospin of baryon $i$.
%
%, {\it e.g.} ${t_{3}}_i=\pm 1/2$ for $i=p,n$.
%

The thermodynamic potential density for baryons in the formalism of RMF is given by
\begin{eqnarray}
\Omega_{B}&=&-T\sum_{i}2\int\frac{d^{3}\bm{k}}{(2\pi)^{3}}\left\{\ln\left[1+e^{-(E^{*}_{i}(k)-\mu^{*}_{i})/T}\right]\right.\nonumber\\
&& \left. +\ln\left[1+e^{-(E^{*}_{i}(k)+\mu^{*}_{i})/T}\right]\right\} \nonumber \\
&&+\frac{1}{2}m_{\sigma}^{2}\sigma^{2} + \frac{1}{3}b\,m_{N}^{} ( {g_{\sigma N}^{}} \sigma)^{3}\nonumber\\
&&+\frac{1}{4}c( {g_{\sigma N}^{}} \sigma)^{4} - \frac{1}{2}m^{2}_{\omega} \omega_{0}^{2} - \frac{1}{2}m^{2}_{\rho} \rho_{03}^{2}\,,
\label{eq:thermo}
\end{eqnarray}
with temperature $T$ and $E^{*}_{i}=(p^{2}+{m^{*}_{i}}^{2})^{1/2}$. $\mu^{*}_{i}$ is the effective chemical potential, which is defined as
\begin{eqnarray}
\mu^{*}_{i}=\mu_{i} - {g_{\omega i}^{}} \omega_{0}-{t_{3}}_{i} {g_{\rho i}^{}} {\rho_{03}^{}} \,.\label{eq:chemi}
\end{eqnarray}
Employing Eq.~(\ref{eq:thermo}), one can easily obtain other thermodynamic quantities, the number density, for instance, reading
\begin{eqnarray}
\rho_{i}^{} &=&-\frac{\partial \Omega_{B}}{\partial {\mu_{i}^{}} }\nonumber\\
&=&2\int\frac{d^{3}\bm{k}}{(2\pi)^{3}} \Big{[} f(E^{*}_{i}(k))-\bar{f}(E^{*}_{i}(k)) \Big{]} \,,\label{eq:density}
\end{eqnarray}
with the Fermi-Dirac distribution functions being
\begin{eqnarray}
f(E^{*}_{i}(k))&=&\frac{1}{\exp[(E^{*}_{i}(k)-\mu^{*}_{i})/T]+1}\,,\label{eq:fermi}\\
\bar{f}(E^{*}_{i}(k))&=&\frac{1}{\exp[(E^{*}_{i}(k)+\mu^{*}_{i})/T]+1} \, . \label{eq:antifermi}
\end{eqnarray}
And the entropy density is given by
\begin{eqnarray}
S_{B}&&=-\frac{\partial \Omega_{B}}{\partial T}\nonumber\\
=&&\frac{2}{T} \int  \frac{d^{3}\bm{k}}{(2\pi)^{3}}  \nonumber\\
&&\times\sum_{i} \left[ \left( E^{*}_{i}(k)+\frac{k}{3}\frac{d E^{*}_{i}(k)}{dk}-\mu^{*}_{i}\right) f(E^{*}_{i}(k))\right.\nonumber\\
&&+\left.\left( E^{*}_{i}(k)+\frac{k}{3}\frac{d E^{*}_{i}(k)}{d k}+\mu^{*}_{i}\right) \bar{f}(E^{*}_{i}(k)) \right] \, .
\label{eq:entropy}
\end{eqnarray}
The pressure is given by
\begin{eqnarray}
p_{B}^{}& =&-\Omega_{B}\nonumber\\
&=&\sum_{i}2\int\frac{d^{3}\bm{k}}{(2\pi)^{3}} \frac{k}{3}\frac{d E^{*}_{i}(k)}{dk} \Big{[} f(E^{*}_{i}(k))+\bar{f}(E^{*}_{i}(k)) \Big{]} \nonumber\\
&&-\frac{1}{2}m_{\sigma}^{2}\sigma^{2}-\frac{1}{3}b\,m_{N}( {g_{\sigma N}^{}} \sigma)^{3} -\frac{1}{4}c({g_{\sigma N}^{}} \sigma)^{4}\nonumber\\
&&+\frac{1}{2}m^{2}_{\omega} \omega_{0}^{2}+\frac{1}{2}m^{2}_{\rho} \rho_{03}^{2}\,. \label{eq:pressure}
\end{eqnarray}
Moreover, the energy density for the strongly interacting baryons takes the form:
\begin{eqnarray}
\varepsilon_{B}^{} &= &TS_{B}+\sum_{i} {\mu_{i}^{}} {\rho_{i}^{}} - {p_{N}^{}} \nonumber\\
&=&\sum_{i}2\int\frac{d^{3}\bm{k}}{(2\pi)^{3}} E^{*}_{i}(k)\Big{[} f(E^{*}_{i}(k))+\bar{f}(E^{*}_{i}(k)) \Big{]} \nonumber\\
&&+\frac{1}{2}m_{\sigma}^{2}\sigma^{2}+\frac{1}{3}b\,m_{N}({g_{\sigma N}^{}} \sigma)^{3} +\frac{1}{4}c({g_{\sigma N}^{}} \sigma)^{4}\nonumber\\
&&+\frac{1}{2}m^{2}_{\omega} \omega_{0}^{2}+\frac{1}{2}m^{2}_{\rho} \rho_{03}^{2}\, \label{eq:energy}.
\end{eqnarray}

In the mean field approximation, expected values of the meson fields are determined through their stationary conditions, {\it i.e.},
\begin{eqnarray}
\frac{\partial \Omega_{B}}{\partial \sigma}=\frac{\partial
\Omega_{B}}{\partial \omega_{0}^{}}=\frac{\partial
\Omega_{B}}{\partial \rho_{03}^{}}=0,
\end{eqnarray}
which produce the following three equations of motion for mesons:
\begin{eqnarray}
m_{\sigma}^{2} \sigma & = &\sum_{i}g_{\sigma i}\rho^{s}_{i}-b\,m_{N}{g_{\sigma N}^{3}}\sigma^{2}-c{g_{\sigma N}^{4}}\sigma^{3} \,, \label{eq:eommeson1}\\
{m_{\omega}^{2}} \omega_{0}^{} &= &\sum_{i}g_{\omega i}\rho_{i}^{}\,, \label{eq:eommeson2}\\
{m_{\rho}^{2}} {\rho_{03}^{}} & = &\sum_{i}g_{\rho i}t_{3i}^{} \rho_{i}^{} \,, \label{eq:eommeson3}
\end{eqnarray}
with the scalar density for baryons given by
\begin{eqnarray}
\rho^{s}_{i}=2\int\frac{d^{3}\bm{k}}{(2\pi)^{3}}\frac{m^{*}_{i}}{E^{*}_{i}(k)} \Big{[} f(E^{*}_{i}(k))+\bar{f}(E^{*}_{i}(k)) \Big{]} \,.
\label{eq:scalardensity}
\end{eqnarray}

Now, it is left to specify the parameters in the RMF, which are fixed by fitting the properties of nuclear matter,
including the saturation baryon number density $\rho_{0}^{}=0.16\:\mathrm{fm^{-3}}$,
binding energy per nucleon $E/A=-16\:\mathrm{MeV}$, effective mass of nucleon at $\rho_{0}^{}$  $m_{N}^{*}=0.75\,m_{N}$ with $m_{N}=938\:\mathrm{MeV}$,
the incompressibility $K=240\:\mathrm{MeV}$, and the symmetry energy $E_{s}=30.5\:\mathrm{MeV}$.
The obtained parameters are given by $({g_{\sigma N}^{}}/m_{\sigma})^{2}\!=\!10.32\:\mathrm{fm^{2}}$,
$({g_{\omega N}^{}}/m_{\omega})^{2}\!=\!5.41\:\mathrm{fm^{2}}$, $({g_{\rho N}^{}}/m_{\rho})^{2}\!=\!3.84\:\mathrm{fm^{2}}$, $b\!=\!6.97\times10^{-3}$ and $c\!=\!-4.85\times10^{-3}$, respectively.
For the couplings between hyperons and mesons, we adopt the following relations: $g_{\sigma H}=0.7g_{\sigma N}$, $g_{\omega H}=0.783g_{\omega N}$ and $g_{\rho H}=0.783g_{\rho N}$,
for more details, see {\it e.g.}~\citep{Glendenning:2000}.

Apart from baryons, the electrons, neutrinos and photons are also relevant in PNS.
In our calculation, we assume them to be non-interacting.
Furthermore, masses of electrons and neutrinos are also neglected. For the electrons, one has
\begin{eqnarray}
\rho_{e}^{} & =&\frac{1}{3\pi^{2}}\left(\mu_{e}^{3} + \pi^{2} {\mu_{e}^{}} T^{2}\right)\,,\\
\varepsilon_{e} & =&\frac{1}{4\pi^{2}}\left(\mu_{e}^{4}+2\pi^{2}\mu_{e}^{2}T^{2}+\frac{7\pi^{4}}{15}T^{4}\right)\,,\\
p_{e}^{} & =&\frac{1}{12\pi^{2}}\left(\mu_{e}^{4}+2\pi^{2}\mu_{e}^{2}T^{2}+\frac{7\pi^{4}}{15}T^{4}\right)\,,\\
S_{e} & =&\frac{T}{3}\left(\mu_{e}^{2}+\frac{7\pi^{2}}{15}T^{2}\right)\,,
\end{eqnarray}
which are the number density, energy density, pressure, and the entropy density, respectively, and $\mu_{e}^{}$ is the electron chemical potential.
For the electron neutrinos, one has
\begin{eqnarray}
\rho_{\nu_{e}}^{} & = &\frac{1}{6\pi^{2}}\left(\mu_{\nu_{e}}^{3}+\pi^{2}\mu_{\nu_{e}}T^{2}\right)\,, \\
\varepsilon_{\nu_{e}}^{} & = &\frac{1}{8\pi^{2}}\left(\mu_{\nu_{e}}^{4}+2\pi^{2}\mu_{\nu_{e}}^{2}T^{2}+\frac{7\pi^{4}}{15}T^{4}\right)\,, \\
p_{\nu_{e}}^{} & = &\frac{1}{24\pi^{2}}\left(\mu_{\nu_{e}}^{4}+2\pi^{2}\mu_{\nu_{e}}^{2}T^{2}+\frac{7\pi^{4}}{15}T^{4}\right)\,, \\
S_{\nu_{e}}^{} & = &\frac{T}{6}\left(\mu_{\nu_{e}}^{2}+\frac{7\pi^{2}}{15}T^{2}\right)  \, ,
\end{eqnarray}
with the chemical potential for electron neutrinos $\mu_{\nu_{e}}^{}$. Since the chemical potentials for $\mu$ and $\tau$ neutrinos are vanishing,
we will not distinguish between them, and use $\nu_{x}^{}$ to denote them. The thermodynamic quantities relevant to $\nu_{x}^{}$ read
\begin{eqnarray}
\varepsilon_{\nu_{x}}= \frac{7\pi^{2}}{60}T^{4},\quad p_{\nu_{x}}=\frac{7\pi^{2}}{180}T^{4},\quad S_{\nu_{x}}  =  \frac{7\pi^{2}}{45}T^{3}\,.
\end{eqnarray}
At last, for photons we have
\begin{eqnarray}
\varepsilon_{\gamma}=\frac{\pi^{2}}{15}T^{4}, \quad
p_{\gamma}^{}=\frac{\pi^{2}}{45}T^{4},   \quad
S_{\gamma}=\frac{4\pi^{2}}{45}T^{3} \, .
\end{eqnarray}

Summing up all the contributions of the components, we obtain the energy density, pressure, and the entropy density of the stellar matter of a newly born NS, as given by
\begin{eqnarray}
\varepsilon & =\varepsilon_{B}^{} +\varepsilon_{e}^{} +\varepsilon_{\nu_{e}}^{} +\varepsilon_{\nu_{x}}^{} +\varepsilon_{\gamma}^{}  \, ,\\
p & = p_{B}^{} + p_{e}^{} + p_{\nu_{e}}^{} +p_{\nu_{x}}^{} +p_{\gamma}^{} \,,\\
S' & =S_{B}^{} + S_{e}^{} +S_{\nu_{e}}^{} +S_{\nu_{x}}^{} +S_{\gamma}^{} \,,\label{eq:entropy2}
\end{eqnarray}
respectively, where we save notation $S$ for other use in the following.
Because of the high temperature and density inside a newly born NS, the timescale of arriving at a beta equilibrium is of order of $\sim 10^{-8}\,\mathrm{s}$~\citep{Fu:2008zzg},
being much smaller than the period of the nonradial oscillations about $10^{-4}\sim 10^{-2}\,\mathrm{s}$.
Thus, it is reasonable to assume that the stellar matter is always in beta equilibrium, which entails that the chemical potentials for the baryon octets read
\begin{eqnarray}
\mu_{i}=\mu_{n}-Q_i(\mu_{e}-\mu_{\nu_e})\, ,\label{eq:beta}
\end{eqnarray}
with electric charge $Q_{i}^{}$ for baryon $i$. Therefore, we have three independent chemical potentials, {\it i.e.}, $\mu_{n}^{}$, $\mu_{e}^{}$, and $\mu_{\nu_{e}}^{}$.
Given a baryon number density $\rho_{B}^{}=\sum_{i} \rho_{i}^{} $, an electron abundance $Y_{e}^{}$ or a lepton abundance $Y_{L}^{}$,
and an entropy per baryon $S=S'/{\rho_{B}^{}}$, combined with the constraint that the stellar matter is electric charge free, {\it i.e.},
$\sum_{j}Q_{j}^{} {\rho_{j}^{}}=0$ with $j$ running over baryons and leptons,
one can obtain the three independent chemical potentials and all other thermodynamic quantities mentioned above, by solving Eqs.~(\ref{eq:eommeson1})--(\ref{eq:eommeson3}) and (\ref{eq:entropy}).

\subsection{\label{sec:MIT}Newly Born Strange Quark Stars---in MIT Bag Model}

In this subsection we employ the simplest model of the strange quark matter, {\it i.e.}, the MIT bag model~\citep{Chodos:1974je,Farhi:1984qu}, to describe the EoS of the matter in a newly born SQS.
A more sophisticated Nambu--Jona-Lasinio model description, which incorporate the information of the dynamical chiral symmetry breaking of QCD, will be delayed to Sec.~\ref{sec:NJL}.
Newly born SQS matter is usually assumed to be composed of three flavor quarks ($u$, $d$, and $s$), electrons, neutrinos, thermal photons and gluons.
Parameters in the bag model are chosen to be $m_{u}=m_{d}=0$, $m_{s}=150\,\mathrm{MeV}$, and a bag constant $B^{1/4}=154.5\,\mathrm{MeV}$,
which, for the SQS matter, produce a mass per baryon of $928\,\mathrm{MeV}$, being slightly smaller than that $931\,\mathrm{MeV}$ in the stablest nucleus ${^{56}}\mathrm{Fe}$,
that is consistent with the conjecture on the strange quark matter~\citep{Witten:1984rs}. For more discussions about the MIT bag model, see, {\it e.g.}, ~\citep{Farhi:1984qu}.

To proceed, we present some thermodynamic quantities relevant to the matter of newly born SQSs. For the $u$ and $d$ quarks, since they are relativistic, {\it i.e.}, massless, one has
\begin{eqnarray}
\rho_{u}^{} & =&\frac{1}{\pi^{2}}\left(\mu_{u}^{3}+\pi^{2} {\mu_{u}^{}} T^{2}\right)\,,\\
\varepsilon_{u} & =& \frac{3}{4\pi^{2}}\left(\mu_{u}^{4}+2\pi^{2}\mu_{u}^{2}T^{2}+\frac{7\pi^{4}}{15}T^{4}\right)\,,\\
p_{u}^{} & =& \frac{1}{4\pi^{2}}\left(\mu_{u}^{4}+2\pi^{2}\mu_{u}^{2}T^{2}+\frac{7\pi^{4}}{15}T^{4}\right)\,,\\
S_{u} & =  &T\left(\mu_{u}^{2}+\frac{7\pi^{2}}{15}T^{2}\right)\, .
\end{eqnarray}
Same expressions also apply to $d$ quarks.
Notations in the equations above are the same as those in Sec.~\ref{sec:NSs}. For the massive strange quark, we have
\begin{eqnarray}
\rho_{s}^{} & = &  6\int\frac{d^{3}\bm{k}}{(2\pi)^{3}}\big{[} f(E_{s}(k))-\bar{f}(E_{s}(k)) \big{]} \,,\\
\varepsilon_{s}^{} & =  &6\int\frac{d^{3}\bm{k}}{(2\pi)^{3}}E_{s}(k) \big{[} f(E_{s}(k))+\bar{f}(E_{s}(k)) \big{]} \,,\\
p_{s}^{} & = & 6\int\frac{d^{3}\bm{k}}{(2\pi)^{3}}\frac{k}{3}\frac{d E_{s}(k)}{d k} \big{[} f(E_{s}(k))+\bar{f}(E_{s}(k)) \big{]} \,,\\
S_{s} & = & \frac{6}{T}\int\frac{d^{3}\bm{k}}{(2\pi)^{3}}\Big{[} \Big{(} E_{s}(k)+\frac{k}{3}\frac{d E_{s}(k)}{d k}-\mu_{s} \Big{)} f(E_{s}(k))\nonumber\\
&& \quad +\Big{(} E_{s}(k)+\frac{k}{3}\frac{d E_{s}(k)}{d k} + {\mu_{s}^{}} \Big{)} \bar{f}(E_{s}(k)) \Big{]}   \, ,
\end{eqnarray}
with $E_{s}(k)=(k^{2}+m_{s}^{2})^{1/2}$ and the fermionic distribution function given in Eqs.~(\ref{eq:fermi}) and (\ref{eq:antifermi}), but with the chemical potential replaced by that of strange quarks.
Thermodynamics relevant to electrons, neutrinos, and thermal photons have already presented in the last subsection, and here what we need in addition are those related to gluons, which read
\begin{eqnarray}
\varepsilon_{g}=\frac{8\pi^{2}}{15}T^{4},   \quad
p_{g}^{} =\frac{8\pi^{2}}{45}T^{4},     \quad
S_{g}=\frac{32\pi^{2}}{45}T^{3},
\end{eqnarray}
where interactions among gluons are neglected as well. Finally, one obtains the thermodynamic quantities describing the stellar matter of a newly born SQS as
\begin{eqnarray}
\varepsilon &=&\Big(\sum_{i=u,d,s}\varepsilon_{i}\Big)+\varepsilon_{e}+\varepsilon_{\nu_{e}}
+\varepsilon_{\nu_{x}}+\varepsilon_{\gamma}+\varepsilon_{g}+B  ,\quad  \\
p & =&\Big(\sum_{i=u,d,s}{p_{i}^{}}\Big) + p_{e}^{} +p_{\nu_{e}}^{} +p_{\nu_{x}}^{} +p_{\gamma}^{} +p_{g}^{} - B    ,\quad\\
S^{\prime} & =&\Big(\sum_{i=u,d,s}S_{i}\Big)+S_{e}+S_{\nu_{e}}+S_{\nu_{x}}+S_{\gamma}+S_{g},\quad\label{eq:SMIT}
\end{eqnarray}
where $B$ is the bag constant. Similar with the stellar matter of a newly born NS, strange quark matter in a newly born SQS is also in beta equilibrium, leading to the relations for the chemical potentials:
\begin{eqnarray}
\mu_{e}^{} +\mu_{u}^{} & =&  \mu_{d}^{} + \mu_{\nu_{e}}^{}  \,  ,  \label{eq:betaMIT1}\\
\mu_{s}^{} & = &\mu_{d}^{}  \,.\label{eq:betaMIT2}
\end{eqnarray}
Like in the NS case, in order to determine the equilibrium configuration of a SQS, one also needs several conservation conditions, such as the baryon number $(\rho_{u}^{} +\rho_{d}^{} +\rho_{s}^{})/3=\rho_{B}^{}$,
lepton number density abundance $Y_{e}^{}+Y_{\nu_{e}}^{}=Y_{L}^{}$, and the electric charge neutral condition
\begin{eqnarray}
\frac{2}{3}\rho_{u}^{} - \frac{1}{3}\rho_{d}^{} - \frac{1}{3}\rho_{s}^{} -\rho_{e}^{} =0
\, .  \label{eq:neutralMIT}
\end{eqnarray}

\subsection{\label{sec:numerical}Nonradial Oscillation Frequencies}

As we have discussed above, neutrinos are trapped in a newly born NS or SQS, and thus the adiabatic sound speed in Eq.~(\ref{eq:cs}) can be rewritten as
\begin{eqnarray}
c_{s}^{2}=\left(\frac{\partial p}{\partial \rho}\right)_{S,\,Y_{\mathrm{L}}}   \,  .   \label{eq:cs2}
\end{eqnarray}
It might have been noticed that, to determine $c_{s}$ as well as the equilibrium sound speed $c_{e}$ in Eq.~(\ref{eq:ce}), equilibrium configuration of a star, {\it i.e.},
the dependence of $\rho_{B}^{}$, $S$, and $Y_{L}$ or $Y_{e}$ on the radial coordinate, has to be provided.
Following ~\citet{Fu:2008bu}, we employ the equilibrium configurations of newly born NSs, obtained in 2D hydrodynamic simulations of core-collapse supernovae by the Arizona Group~\citep{Dessart:2005ck}.
In our calculations we choose three representative instants of time, {\it i.e.}, $t=100$, $200$ and $300\,\mathrm{ms}$, after the core bounces in the core-collapse supernovae.
Since the eigenfrequencies of $g$-mode vary only a little during the first second after the core bounces, for instance,
it has been found that the eigenfrequencies resides in a narrow range of $727\,\mathrm{Hz}\sim 819\,\mathrm{Hz}$, from $0.3\,\mathrm{s}$ to $1\,\mathrm{s}$ in a newly born NS model~\citep{Ferrari:2002ut},
we restrict computations in this work within $300\,\mathrm{ms}$ for simplicity.

We integrate Eqs.~(\ref{eq:Osc1}) and (\ref{eq:Osc2}) from the center to a radius of about $20\,\mathrm{km}$, where convective instabilities set in \citet{Dessart:2005ck}.
The stellar mass of the newly born NS inside the radius of $20\,\mathrm{km}$ is chosen to be $0.8\,\mathrm{M}_{\odot}$, $0.95\,\mathrm{M}_{\odot}$ and $1.05\,\mathrm{M}_{\odot}$
at $t\!=\!100$, $200$ and $300\,\mathrm{ms}$, respectively, in agreement with the supernovae simulations~\citep{Dessart:2005ck}.
The mass increases with time, because mantle materials of the progenitor star are accreted onto the newly born NS continuously.

For newly born SQSs, the thermal profile is unknown.
Therefore, following \citet{Fu:2008bu},
we assume that the dependence of the entropy $S$ and lepton abundance $Y_{L}$ on the radial coordinate for newly born SQSs,
is the same as for NSs.
There have been calculations on the oscillation frequency with completely different profiles~\citep{Wei:2018tts,Flores:2013yqa},
and the result is qualitatively in accordance with ours.
Therefore, the effect of thermal profile should be subleading and our assumption is reasonable.
In comparison with NSs, a newly born SQS has a smaller size, with a radius of about $10\,\mathrm{km}$ where the pressure vanishes.
The stellar mass of the SQS in our calculations is chosen to be the same as the NS, for the three different instants after the core bounces.

\begin{table}[!htbp]
\begin{center}
\caption{Eigenfrequencies (Hz) of quadrupole ($l=2$) oscillations of $g$-mode for newly born NSs and SQSs at three instants of time (ms) after the core bounces (Quoted from Table~I in \citet{Fu:2008bu}).}
\label{tab:gmode}
\begin{tabular}{c|ccc|ccc}
\hline \hline                    %\vspace{0.1cm}
             &\multicolumn{3}{c|}{NS} & \multicolumn{3}{c}{SQS}  \\\hline
radial             &\multicolumn{3}{c|}{time(ms)} & \multicolumn{3}{c}{time(ms)}  \\
order             & $100$ & $200$       & $300$  & $100$ & $200$ & $300$\\
\hline
 $n=1$       & 717.6       & 774.6             & 780.3        &  82.3       & 78.0        & 63.1 \\
 $n=2$       & 443.5       & 467.3             & 464.2        &  52.6       & 45.5        & 40.0   \\
 $n=3$       & 323.8       & 339.0             & 337.5        &  35.3       & 30.8        & 27.8   \\
\hline \hline
\end{tabular}
\end{center}
\end{table}

In this work, for newly born NSs, we extend our former calculations in \citet{Fu:2008bu} to include the hyperon degrees of freedom.
However, since the stellar masses are all just about one solar mass for the three instants of time,
which results in that the baryon densities in the stellar center are less than $2\rho_{0}^{}$,
we find then that the hyperons do not appear in the newly born NSs for all the three instants.
Thus, the eigenfrequencies of the $g$-mode oscillations for newly born NSs calculated in this work are identical to those presented in the Table I in \citet{Fu:2008bu}.
We quote the result here in Table~\ref{tab:gmode} for the convenience of discussion.

Note that $g$-mode quadrupole pulsations of compact stars with low radial orders, especially $n=1$,
are significantly potential sources of gravitational waves~\citep{Owen:2005fn,Lai:2006pr},
which are hopefully to be detected in the near future by such as Advanced LIGO.

One can see from Table.~\ref{tab:gmode} that, for the $g$-mode quadrupole oscillations of NSs with $n=1$,
the eigenfrequencies are $717.6$, $774.6$ and $780.3\,\mathrm{Hz}$, respectively, for the three different instants.
These results are consistent with supernovae simulations~\citep{Burrows:2006ci,Ott:2006qp} as well as those computed from general relativity~\citep{Ferrari:2002ut}.

In contrast to the case of NSs, eigenfrequencies of the $g$-mode for SQSs with $n=1$, $l=2$,
at the three representative instants after the core bounces,
are $82.3$, $78.0$ and $63.1\,\mathrm{Hz}$ respectively, which are lower by an order of magnitude.

Table~\ref{tab:gmode} also shows the results for higher radial orders with $n>1$.
One can still find that eigenfrequencies of NSs are larger than those of SQSs by an order of magnitude.
It was found in ~\citep{Fu:2008bu} that this difference arises from the fact that nucleons,
the major components of the stellar matter of NSs, are massive and non-relativistic, whereas SQSs are composed of relativistic particles.
In fact, non-vanishing $g$-mode eigenfrequencies of a SQS are attributed to a finite strange quark mass $m_s$, since the masses of $u$ and $d$ quarks are negligible.

\begin{figure}[!t]
\begin{center}
\includegraphics[scale=0.6]{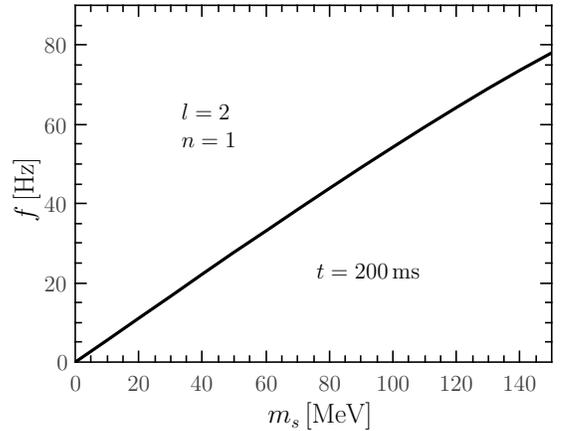}
%\vspace*{-1mm}
\caption{Eigenfrequency of the $g$-mode for a SQS with $l=2$ and $n=1$, as a function of the strange quark mass $m_s$, where we choose $t\!=\!
200\mathrm{ms}$ illustratively.}\label{fig:fms}
\end{center}
\end{figure}

Therefore, if we deal with $m_{s}^{}$ as a parameter, and reduce it from a finite value to zero artificially,
one would have expected that the $g$-mode eigenfrequency of a SQS decreases with $m_{s}^{}$, and finally vanishes when $m_{s}^{}=0$.
This expectation is confirmed by the numerical result shown in Fig.~\ref{fig:fms}, where we have chosen the instant $t\!=\!200\mathrm{ms}$ for an illustrative purpose.
Note that the mass of strange quark is much smaller than that of nucleons, which is the reason why the eigenfrequencies of $g$-mode oscillations of a SQS are much smaller than those of a NS.

\begin{figure*}[t]
\includegraphics[scale=0.44]{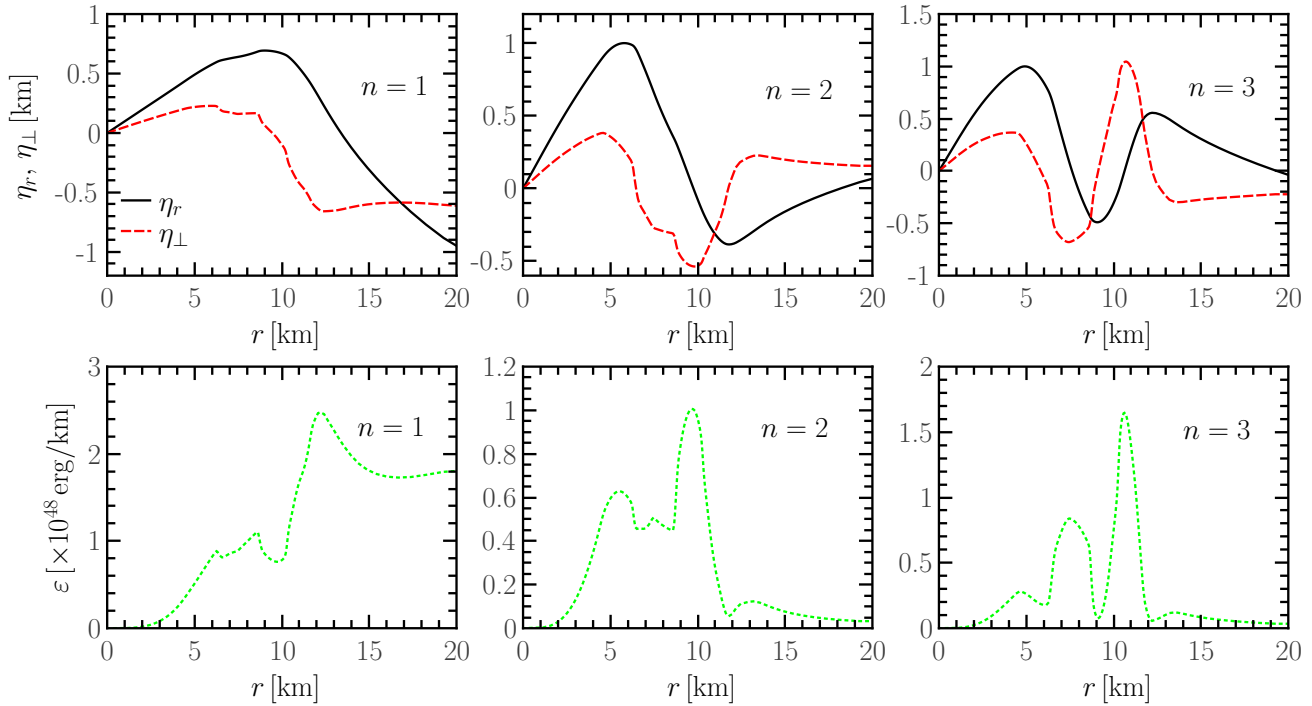}
%\vspace*{-2mm}
\caption{Upper panels: radial $\eta_{r}^{}$ and nonradial $\eta_{\bot}^{}$ components of the displacement in Eq.~(\ref{eq:eta}) as functions of the radial coordinate $r$,
for the $g$-mode oscillations of a newly born NS with $l=2$, and $n=1,\,2,\,3$, respectively.
$t\!=\!200\,\mathrm{ms}$ is chosen as a representative instant of time. Lower panels: corresponding energy of stellar oscillations per unit of the radial coordinate as a function of $r$.}\label{fig:waveformgNS}
\end{figure*}

\begin{figure*}[t]
\begin{center}
\includegraphics[scale=0.44]{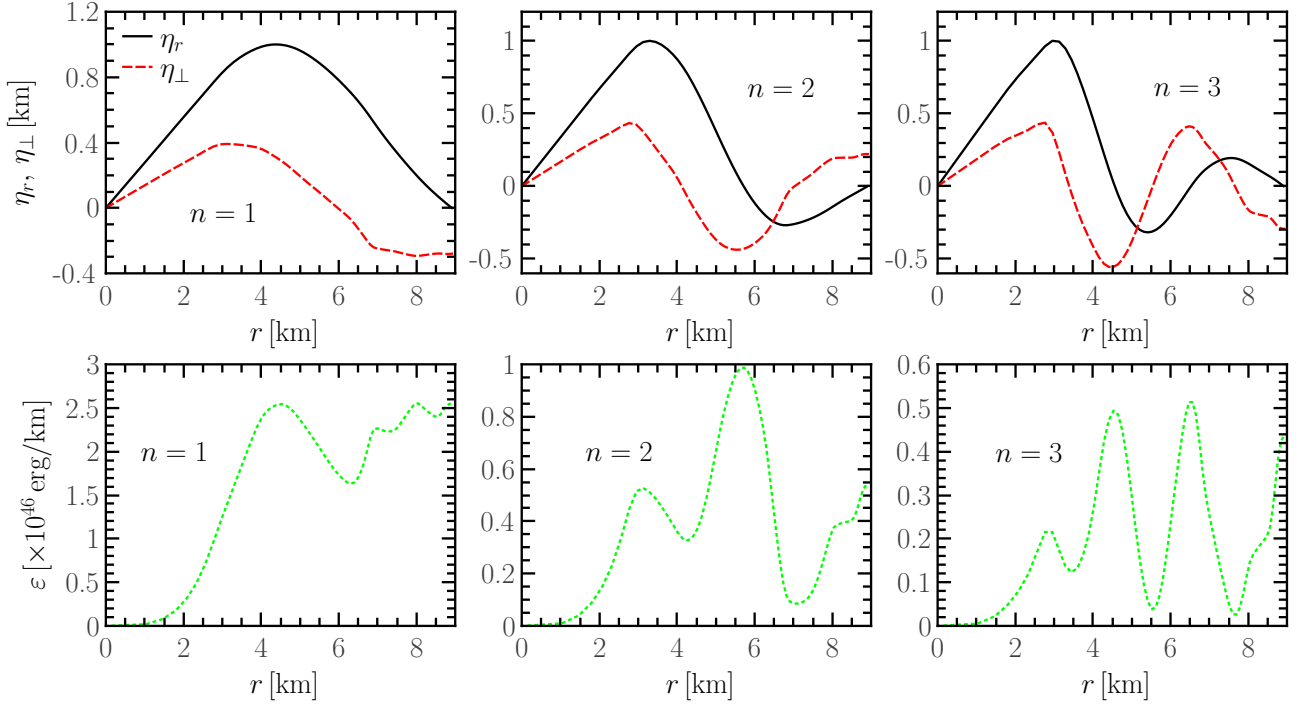}
%\vspace*{-2mm}
\caption{Same as Fig.~\ref{fig:waveformgNS}, but for a newly born SQS.}\label{fig:waveformgSQS}
\end{center}
\end{figure*}

In Fig.~\ref{fig:waveformgNS} and Fig.~\ref{fig:waveformgSQS} we show the eigenfunctions of the quadrupole oscillations of $g$-mode,
described by the components of the displacement $\eta_{r}^{}$ and $\eta_{\bot}^{}$ as functions of the radial coordinate,
for the newly born NS and SQS, respectively.
Here we choose the radial order $n=1,\,2,\,3$, and $t\!=\!200\,\mathrm{ms}$ after the core bounce as a representative instant.
Note that the amplitude of oscillations is normalized so that the maximal value of $\eta_{r}^{}$ is $1\,\mathrm{km}$.

Furthermore, the energy of the stellar oscillation of an eigenmode is given by \citet{Reisenegger:1992APJ}
\begin{equation}\label{eqn:ModeEnergy}
E = \frac{\omega^{2}}{2}\int_{0}^{R}\rho r^{2} \big{[} \eta_{r}^{2} + l(l+1)\eta_{\bot}^{2} \big{]} d r\, . %\label{mode_E}
\end{equation}

In Fig.~\ref{fig:waveformgNS} and Fig.~\ref{fig:waveformgSQS} we also show the oscillation energy per unit of the radial coordinate, viz.
\begin{equation}
\varepsilon=\frac{\omega^{2}}{2}\rho
r^{2} \big{[} \eta_{r}^{2}+l(l+1)\eta_{\bot}^{2} \big{]} \, ,\label{E_density}
\end{equation}
for the $g$-mode oscillations of the newly born NS and SQS, respectively.
We find that the energy density $\varepsilon$ in \Eq{E_density} for the SQS is lower than that for the NS by two orders of magnitude, because of the big difference of eigenfrequencies.

\begin{figure*}[!tp]
\includegraphics[scale=0.44]{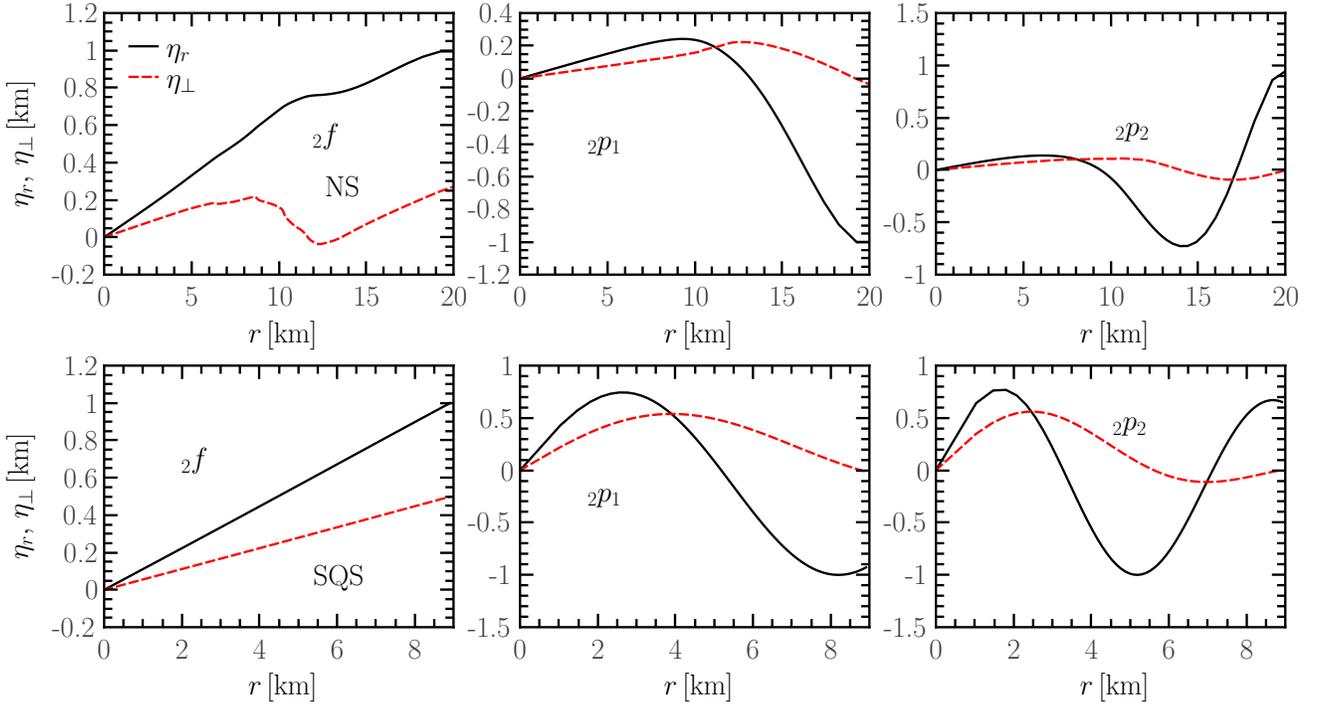}
%\vspace*{-3mm}
\caption{$\eta_{r}^{}$ and $\eta_{\bot}^{}$ as functions of the radial coordinate,
	for the $f$- and $p$-modes of a newly born NS (upper panels) and SQS (lower panels) with $l=2$. $t\!=\!200\,\mathrm{ms}$ is chosen as a representative instant of time.}\label{fig:waveformfp}
\end{figure*}

Up to now, we have only discussed the stellar nonradial oscillations of $g$-mode.
Besides the gravitational mode, there are other modes for nonradial oscillations, such as the $f$-mode, {\it i.e.},
the mode with the radial order $n=0$, and the $p$-mode, both of which can also be obtained by solving the nonradial oscillation equations in Eq.~(\ref{eq:Osc1}) and Eq.~(\ref{eq:Osc2}).
In this work we would like to present some results for the $f$- and $p$-modes, in comparison with the relevant ones of $g$-mode.
In the following we employ $_{l}p_{n}$ to denote the pressure mode with angular index $l$ and the radial order $n$, and $_{l}f$ the fundamental mode with angular index $l$.

\begin{table}[!tp]
\begin{center}
\caption{Eigenfrequencies (Hz) of quadrupole ($l=2$) oscillations of $f$- and $p$-modes for newly born NSs and SQSs at three instants of time (ms) after the core bounces.}
\label{tab:fpmode}
\begin{tabular}{c|ccc|ccc}
\hline \hline                    %\vspace{0.1cm}
      &\multicolumn{3}{c|}{NS} & \multicolumn{3}{c}{SQS}  \\
\hline
     &\multicolumn{3}{c|}{time(ms)} & \multicolumn{3}{c}{time(ms)}  \\
Modes& $100$&$200$&$300$&$100$&$200$&$300$\\
\hline
 $_{2}f$     & 1103& 1133 & 1176 &  2980  & 2997  & 3016 \\
 $_{2}p_{1}$ & 2265& 2426 & 2494 &  18282 & 17330 & 16792   \\
 $_{2}p_{2}$ & 3780& 4054 & 4179 &  28792 & 27288 & 26438   \\
 $_{2}p_{3}$ & 5319& 5702 & 5869 &  38988 & 36950 & 35798   \\
\hline \hline
\end{tabular}
\end{center}
\end{table}

In Table~\ref{tab:fpmode} we show our obtained eigenfrequencies of quadrupole oscillations of $f$- and $p$-modes for newly born NSs and SQSs at the three representative instants.
One can observe evidently that the eigenfrequencies of $f$- and $p$-modes are larger than those of the $g$-mode for both NSs and SQSs, which is because they have different origins.
Oscillations of $g$-mode arise from the buoyancy inside the star, but those of $f$- and $p$-modes result from the pressure.
As discussed in detail in Ref.~\citet{Fu:2008bu}, the eigenfrequencies of $g$-mode are closely related with the difference of the equilibrium and adiabatic sound speeds,
{\it i.e.}, the Brunt-V\"{a}is\"{a}l\"{a} frequency in Eq.~(\ref{eq:BV}), while those of $f$- and $p$-modes are connected with the value of sound speeds.
The difference between the two sound speeds are much smaller than each of them, thus the eigenfrequencies of $g$-mode are smaller than those of $f$- and $p$-modes.
In Table~\ref{tab:fpmode} we also find that, for the $f$- and $p$-modes, eigenfrequencies of SQSs are larger than those of NSs, in contradistinction to the $g$-mode.
This is because the SQSs are more compact than NSs, and specifically densities in the outer region of the NSs are significantly low, which leads to smaller sound speeds there.
In Fig.~\ref{fig:waveformfp} we show the eigenfunctions of several $f$- and $p$-modes for the NS and SQS.

\section{\label{sec:NJL}The effect of dynamical chiral symmetry breaking in Strange Quark Stars}

In \sec{sec:MIT} we employ the MIT bag model to describe the quark matter and construct the configuration of the SQS.
Nonperturbative QCD is characteristic of the dynamical chiral symmetry breaking (DCSB) at low energy.
Since there is no explicit demonstration of the DCSB in the bag model, it can not describe the important feature of QCD.
In this section we would like to adopt the Nambu--Jona-Lasinio (NJL) model \citep{Nambu:1961fr,Nambu:1961tp},
which possesses the chiral symmetry and DCSB, for more details about the NJL model and applications of the model in hadron physics and QCD phase diagram, see, {\it e.g.},
\citet{Klevansky:1992qe,Hatsuda:1994pi,Buballa:2003qv}.
The Lagrangian density for the 2+1 flavor NJL model is given by \citet{Rehberg:1995kh}
\begin{eqnarray}
\mathcal{L}&=&\, \bar{\psi}\left(i\gamma_{\mu}\partial^{\mu} \!- \! \hat{m}_{0}\right)\psi \nonumber\\
&&+G\sum_{a=0}^{8}\left[\left(\bar{\psi}\tau_{a}\psi\right)^{2} \! + \! \left(\bar{\psi}i\gamma_{5}\tau_{a}\psi\right)^{2}\right]   \nonumber \\
&&-K\left[\textrm{det}_{f}\left(\bar{\psi}\left(1+\gamma_{5}\right)\psi\right) +\textrm{det}_{f}\left(\bar{\psi}\left(1-\gamma_{5}\right)\psi\right)\right]
 ,\quad\quad\label{eq:NJLlagragian}
\end{eqnarray}
with the quark fields $\psi=(\psi_{u},\psi_{d},\psi_{s})^{T}$, the matrix of the current quark masses $\hat{m}_{0}=\textrm{diag}(m_{u},m_{d},m_{s})$.
We choose $m_{l}^{}:=m_{u}^{}=m_{d}^{}$ and $m_{l}^{}<m_{s}^{}$ as same as the case in \sec{sec:MIT}.
The four fermion interactions in Eq.~(\ref{eq:NJLlagragian}) is symmetric under the transformations of $U(3)_{L}\otimes U(3)_{R}$ in the flavor space,
with the interaction strength $G$. $\tau_{0}^{}=\sqrt{\frac{2}{3}} \mathbf{1}_{f}$ and Gell-Mann matrices $\tau_{i}^{}\,(i=1,\ldots,8)$ are normalized such that we have $\mathrm{tr}(\tau_{a}\tau_{b})=2\delta_{ab}$.
The 't Hooft interactions in Eq.~(\ref{eq:NJLlagragian}), with the coupling strength $K$, break the symmetry of $U_{A}(1)$, but keep that of $SU(3)_{L}\otimes SU(3)_{R}$.
The parameters in the model read: the current mass of light quarks $m_{l}=5.5\;\mathrm{MeV}$, the strange quark mass $m_{s}=140.7\;\mathrm{MeV}$,
coupling strengths $G\Lambda^{2}=1.835$ and $K\Lambda^{5}=12.36$, with the UV cutoff $\Lambda=602.3\;\mathrm{MeV}$.
These parameters are fixed by fitting meson properties in vacuum,
including $\pi$ meson mass $m_{\pi}=135.0\;\mathrm{MeV}$, $K$ meson mass $m_{K}=497.7\;\mathrm{MeV}$,
$\eta^{\prime}$ meson mass $m_{\eta^{\prime}}=957.8\;\mathrm{MeV}$, and the $\pi$ decay constant $f_{\pi}=92.4\;\mathrm{MeV}$, for more details, see, {\it e.g.}, \citet{Rehberg:1995kh}.

The thermodynamic potential density in the mean field approximation reads
\begin{eqnarray}
\Omega_{\mathrm{QM}}&=&-2N_{c}\!\!\sum_{i=u,d,s}\int\frac{d^{3}\bm{k}}{(2\pi)^{3}}\bigg\{\beta^{-1}\ln\Big[1+e^{-\beta(E_{i}^{}(k) - \mu_{i}^{})}\Big]\nonumber\\
&&+\beta^{-1}\ln\Big[1+e^{-\beta(E_{i}^{}(k) + \mu_{i}^{})}\Big]\,+E_{i}^{} \bigg\}\nonumber\\
&&+2G\left({\phi_{u}}^{2} +{\phi_{d}}^{2}+{\phi_{s}}^{2}\right) - 4K \phi_{u}^{}\,\phi_{d}^{}\,\phi_{s}^{}+C\,,
\label{eq:NJLpoten}
\end{eqnarray}
where $C$ is a constant, to be determined in the following.
The subscript $_{\mathrm{QM}}$ denotes quark matter and $N_{c}^{}=3$ is the number of colors.
The dispersion relation for quarks is given by
\begin{equation}\label{eq:NJLdispersion}
E_{i}^{}(k)=\sqrt{k^{2}+M_{i}^{2}}\,,
\end{equation}
with the constituent quark mass of flavor $i$, {\it i.e.}, $M_{i}^{}$ being
\begin{equation}
M_{i}^{}=m_{0}^{i} - 4G\phi_{i}^{} + 2K\phi_{j}^{}\,\phi_{k}^{},\label{eq:constituentmass}
\end{equation}
where $\phi_{i}^{}$ is the quark condensate of flavor $i$, reading
\begin{eqnarray}
\phi_{i}^{} = -2N_{c}\int\frac{d^{3}\bm{k}}{(2\pi)^{3}}\frac{M_{i}}{E_{i}}
\Big{[} 1-f\big(E_{i}(k)\big) - \bar{f}\big(E_{i}(k)\big) \Big{]}, \quad\, \label{eq:condensate}
\end{eqnarray}
where $f$ and $\bar{f}$ are the fermionic distribution functions in Eqs.~(\ref{eq:fermi}) and (\ref{eq:antifermi}) with $E^{*}$ and $\mu^{*}$ replaced with $E$ and $\mu$, respectively. In the same way, employing the thermodynamic potential in Eq.~(\ref{eq:NJLpoten}), one can obtain other thermodynamic quantities, such as the quark number density reading
\begin{eqnarray}
\rho_{i}^{}& =&-\frac{\partial \Omega_{\mathrm{QM}}}{\partial \mu_{i}}\nonumber\\
&=&\, 2N_{c}\int\frac{d^{3}\bm{k}}{(2\pi)^{3}} \Big{[} f\big(E_{i}(k)\big) - \bar{f}\big(E_{i}(k)\big) \Big{]}\,,
\label{eq:NJLdensity}
\end{eqnarray}
the entropy density
\begin{eqnarray}
&S_{\mathrm{QM}}&=-\frac{\partial \Omega_{\mathrm{QM}}}{\partial T}\nonumber\\
\!&\!=\! &\!\!\! \frac{2N_{c}}{T}\sum_{i=u,d,s}\int\frac{d^{3}\bm{k}}{(2\pi)^{3}} \left[ \left(E_{i}(k) +\frac{k^{2}}{3E_{i}(k)}-\mu_{i}\right)\right. \qquad \nonumber\\
\!& & \!\!\!\! \times \! \left. f\!\left(E_{i}(k)\right) \! + \! \left( \! E_{i}(k) \!+ \! \frac{k^{2}}{3E_{i}(k)} \!+ \! \mu_{i}\Big)\bar{f}\big(E_{i}(k)\right)\right] , \quad
\label{eq:NJLentropy}
\end{eqnarray}
the pressure
\begin{eqnarray}
p_{\mathrm{QM}}^{} &= &-\Omega_{\mathrm{QM}}^{}     \nonumber\\
&=&\, 2N_{c}\sum_{i=u,d,s}\int\frac{d^{3}\bm{k}}{(2\pi)^{3}}
\frac{k^{2}}{3E_{i}(k)} \Big{[} f\big(E_{i}(k)\big) \! + \! \bar{f}\big(E_{i}(k)\big) \Big{]} \nonumber\\
&&+2N_{c}\sum_{i=u,d,s}\int\frac{d^{3}\bm{k}}{(2\pi)^{3}}E_{i}(k)\nonumber\\
&&-2G\left({\phi_{u}}^{2}
+{\phi_{d}}^{2}+{\phi_{s}}^{2}\right)+4K\phi_{u}\,\phi_{d}\,\phi_{s}-C\,, \label{eq:NJLpressure}
\end{eqnarray}
from which we can define an effective bag constant as
\begin{eqnarray}
B_{\textrm{eff}}&=&-2N_{c}\sum_{i=u,d,s}\int\frac{d^{3}\bm{k}}{(2\pi)^{3}}E_{i}(k)\nonumber\\
&&+2G\left({\phi_{u}}^{2}
+{\phi_{d}}^{2}+{\phi_{s}}^{2}\right)-4K\phi_{u}\,\phi_{d}\,\phi_{s}+C. \quad\label{eq:Beff}
\end{eqnarray}
In contrast to the bag constant in the MIT bag model, $B_{\textrm{eff}}$ in Eq.~(\ref{eq:Beff}) is dependent on the quark condensates, and thus on the density, rather than an absolute constant. Finally, one obtains the energy density as given by
\begin{eqnarray}
\varepsilon_{\mathrm{QM}}^{}&=&\, TS_{\mathrm{QM}}+\sum_{i=u,d,s}\mu_{i}\rho_{i} -p_{\mathrm{QM}}^{} \nonumber\\
&=& \, 2N_{c}\sum_{i=u,d,s}\int \!\! \frac{d^{3}\bm{k}}{(2\pi)^{3}} E_{i}^{}(k)
\Big{[} f\big(E_{i}^{}(k)\big)\! +\bar{f}\big(E_{i}^{}(k)\big) \Big{]} \nonumber\\
&&+B_{\textrm{eff}}\,. \label{eq:NJLenergy}
\end{eqnarray}
Note that the constant $C$ in Eq.~(\ref{eq:NJLpoten}) is determined with $\varepsilon_{\mathrm{QM}}^{}= -p_{\mathrm{QM}}^{} =0$ in the vacuum.

%%%%%%%%%%%%%%%%%%%%%%%%%%%%%

\begin{figure}[t]
\begin{center}
\includegraphics[scale=0.55]{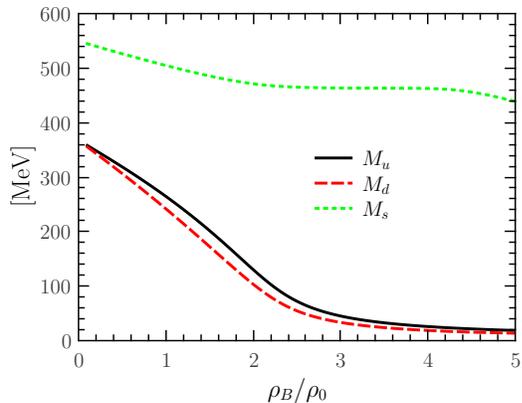}
\vspace*{-1mm}
\caption{Constituent quark masses as functions of the baryon number density, calculated in the NJL model, for the charge-neutral quark matter at zero temperature and neutrino abundance.}\label{fig:quarkmass}
\end{center}
\end{figure}
%%%%%%%%%%%%%%%%%%%%%%%%%

%%%%%%%%%%%%%%%%%%%%%%%%%%%%%
\begin{figure*}[htb]
\begin{center}
\includegraphics[scale=0.6]{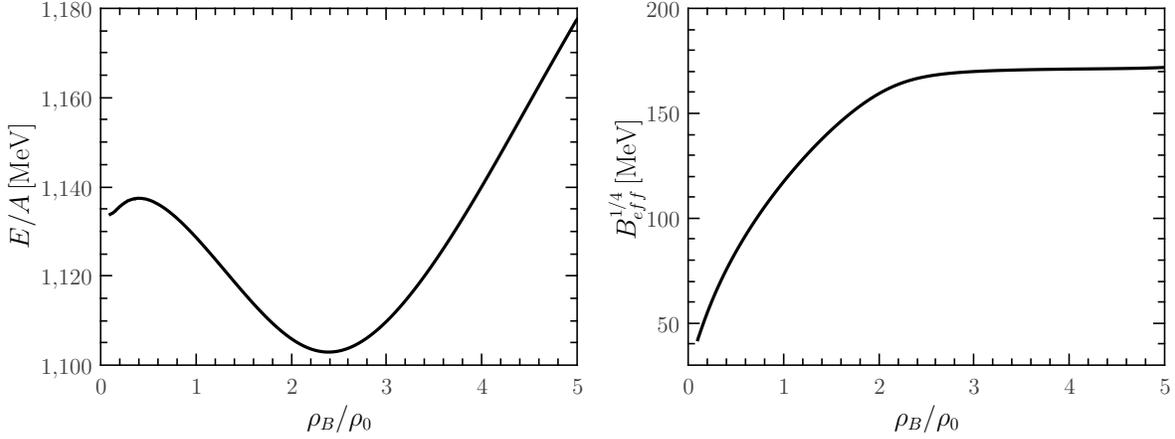}
%\vspace*{-2mm}
\caption{Energy per baryon number $E/A$ (left panel) and the effective bag constant $B_{\textrm{eff}}^{1/4}$ defined in Eq.~(\ref{eq:Beff}) (right panel) as functions of the baryon number density for the charge-neutral quark matter at zero temperature and neutrino abundance.}\label{fig:EAB}
\end{center}
\end{figure*}
%%%%%%%%%%%%%%%%%%%%%%%%%

%%%%%%%%%%%%%%%%%%%%%%%%%%%%%
\begin{figure*}[htb]
\includegraphics[scale=0.6]{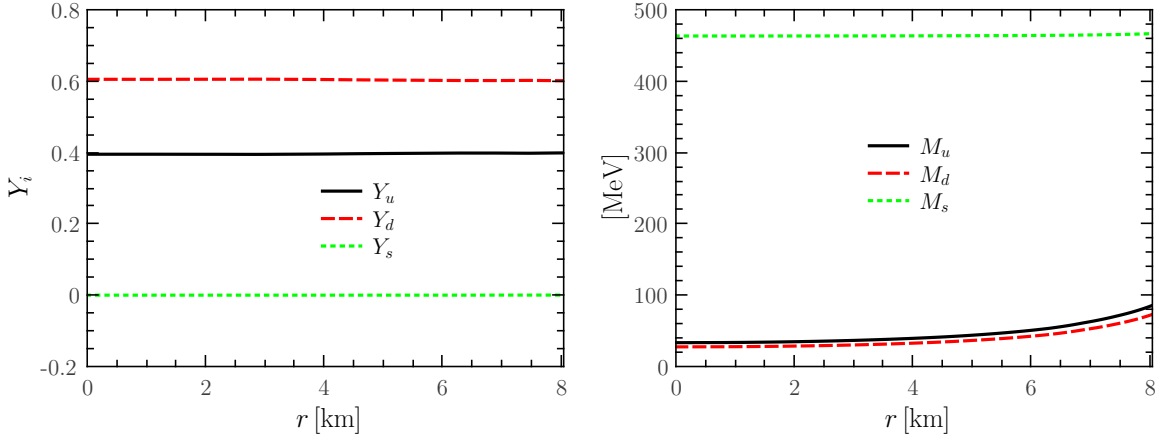}
\vspace*{-1mm}
\caption{Abundances of the three flavor quarks (left panel) and their corresponding constituent quark masses (right panel) as functions of the stellar radial coordinate for a newly born SQS. $t\!=\!200\,\mathrm{ms}$ is chosen as a representative instant of time.}\label{fig:Yimass}
\end{figure*}
%%%%%%%%%%%%%%%%%%%%%%%%%

As we have discussed above, the $g$-mode frequencies of newly born SQSs are dependent on the masses of quarks.
In Fig.~\ref{fig:quarkmass} we show the constituent quark masses of the three flavors as functions of the baryon number density,
in unit of the saturation baryon number density $\rho_{0}^{}$, for the charge-neutral quark matter at zero temperature and neutrino abundance.
We find that with the increase of the density, the dynamical chiral symmetry is restored gradually,
and the constituent quark masses of $u$ and $d$ quarks decrease from $350\,\mathrm{MeV}$ to $50\,\mathrm{MeV}$ when the density increases from $0.1\,\rho_{0}^{}$ to $3\,\rho_{0}^{}$.
Since the current mass of the strange quark is larger, its constituent mass decreases relatively milder in comparison with the light quarks.
Fig.~\ref{fig:EAB} depicts the energy per baryon number $E/A$ and the effective bag constant $B_{\textrm{eff}}^{1/4}$ defined in Eq.~(\ref{eq:Beff}) versus the density for the charge-neutral quark matter at zero temperature.

One can observe easily from the figure that when the charge-neutral quark matter is in equilibrium, where $E/A$ has a minimum value,
$E/A$ is about $1100\,\mathrm{MeV}$, larger than the energy per nucleon in Iron nuclei about $930\,\mathrm{MeV}$.
As we will see in the following, the large constituent mass of the $s$ quark leads to its very low abundance in the charge-neutral quark matter. In turn, the Fermi energy is enhanced and the $E/A$ increases.

Furthermore, we find that the effective bag constant $B_{\textrm{eff}}^{1/4}$ increases from about $40\,\mathrm{MeV}$ to $170\,\mathrm{MeV}$ as the density increases from $0.1\,\rho_{0}^{}$ to $3\,\rho_{0}^{}$.
Such a baryon number density dependence of the $B_{\textrm{eff}}^{1/4}$ is qualitatively consistent with that given in Ref.~\citet{Buballa:2003qv}.

It is left to employ the NJL model to construct the configuration of the SQS, for more discussions about the charge-neutral quark matter, see, {\it e.g.}, \citet{Buballa:1998pr}.
In Fig.~\ref{fig:Yimass} we present the abundances of the three flavor quarks and their corresponding constituent quark masses as functions of the stellar radial coordinate for a newly born SQS.
One can see that $u$ and $d$ quarks in the NJL model are still relativistic, and their constituent quark masses range from about $30\,\mathrm{MeV}$ to $85\,\mathrm{MeV}$ inside the star.
The strange quark in the NJL model, however, is non-relativistic, since its constituent mass is much larger than those of light quarks, as seen in the right panel of Fig.~\ref{fig:Yimass}.
It is the large constituent quark mass, which results in very low abundance of strange quarks inside the SQS.
Therefore, all the constituents of a SQS, described by the NJL model, are relativistic as well, which are similar with the case in the MIT model.
Thus, it is expected that the $g$-mode eigenfrequencies of a SQS, with its EoS described by the NJL model, are also smaller than those of a newly born NS.

\begin{table}[!htbp]
\begin{center}
\caption{
	Eigenfrequencies (Hz) of quadrupole ($l=2$) oscillations  for newly born SQSs, with the quark matter described by the NJL model, at three instants of time (ms) after the core bounces.
	}
\label{tab:NJLgmode}
\begin{tabular}{cccc}
\hline \hline                    \vspace{0.1cm}
Mode  & $t\!=\!100$&$t\!=\!200$&$t\!=\!300$ \\\hline
 $_{2}g_{2}$ & 60.1 & 57.0 & 51.8    \\
 $_{2}g_{1}$ & 100.2& 115.4& 107.4    \\
 $_{2}f$     & 3482 & 3510 & 3530    \\
 $_{2}p_{1}$ & 19477& 18632& 18159  \\
\hline \hline
\end{tabular}
\end{center}
\end{table}

In Table~\ref{tab:NJLgmode} we list some of our obtained oscillation eigenfrequencies for the different modes of newly born SQSs in the NJL model.
In the same way, we choose three representative instants of time after the core bounce.
We find that the $g$-mode eigenfrequencies of newly born SQSs with $l=2$, $n=1$ in the NJL model are $100.2\,\mathrm{Hz}$, $115.4\,\mathrm{Hz}$, $107.4\,\mathrm{Hz}$ at the three instants, respectively.
These values are a bit larger than those of the relevant modes in the MIT bag model as shown in Table~\ref{tab:gmode},
because of the finite constituent masses of light quarks in the NJL model, but they are still much smaller than those of newly born NSs.
The $f$-mode frequency is about $\sim 3500$Hz
and the $p_{1}$-mode is about $\sim 19000$Hz for all the three instant of time,
which is of the same order as those obtained via MIT bag model,
and is significantly larger than the NS results.

\section{In case of Stiff Equation of state}\label{sec:stiffEOS}

For hadron matter, when the Fermi momentum of proton and neutron exceed the mass of hyperons,
the hyperon degree of freedom is likely to appear,
and the equation of state will be softened, which reduces the maximum mass of the neutron star.
For quark matter, the equation of state is generally softer than that of hadron matter,
and hence the maximum mass of strange quark star is even smaller than that of neutron star.
However, several years ago, compact stars with mass of about $2M_{\odot}$ were discovered~\citep{Demorest:2010bx,Antoniadis:2013pzd,Fonseca:2016tux,NANOGrav:2017wvv,Linares:2018ppq,NANOGrav:2019jur},
and such a mass is larger than the predicted values of most models,
which is a great challenge to the theory of dense matter.
There have been different approaches towards this problem,
and in this section we will briefly introduce a stiff hadron model and a stiff quark model, and calculate the frequency of the gravitational wave under stiff EoSs.

\subsection{Extended Relativistic Mean Field Theory}\label{sec:stiffRMF}
For pure hadron matter, if we only include nucleons ($p$ and $n$) in our model,
	the EoS can be stiff enough to support a 2-solar-mass neutron star,
but after the inclusion of hyperon, for most of the models, the maximum mass of neutron star will be reduced below $2M_{\odot}$.
This is the so-called ``hyperon puzzle''.
In \citet{Lopes:2013cpa} the ``hyperon puzzle'' is solved under an extended version of RMF.
The Lagrangian for the model is
\begin{equation}
	\mathcal{L}=\mathcal{L}_{B}+\mathcal{L}_{\textrm{lep}}+\mathcal{L}_{M}+\mathcal{L}_{\textrm{int}},
\end{equation}
where $\mathcal{L}_{B}$ is the Lagrangian for free baryons, $\mathcal{L}_{\textrm{lep}}$ is the Lagrangian for leptons,
$\mathcal{L}_{M}$ is the Lagrangian for mesons, and $\mathcal{L}_{\textrm{int}}$ is the interaction term.

For baryons, the baryon octet are included, and $\mathcal{L}_{B}$ is
\begin{equation}
	\mathcal{L}_{B}=\sum_{i}\bar{\Psi}_{i}\left(i\gamma_{\mu}\partial^{\mu}-m_{i}\right)\Psi_{i},
\end{equation}
where $\Psi_{i}$ ($i=p,n,\Lambda,\Sigma^{\pm,0},\Xi^{-,0}$) are the baryon octet fields.

The interaction between mesons and baryons is
\begin{eqnarray}
\mathcal{L}_{\textrm{int}}&=&\sum_{i}g_{\sigma i}\bar{\Psi}_{i}\sigma\Psi_{i}-g_{\omega i}\bar{\Psi}_{i}\gamma_{\mu}\omega^{\mu}\Psi_{i}\nonumber\\
&&-g_{\phi i}\Psi_{i}\gamma_{\mu}\phi^{\mu}\Psi_{i}-g_{i\rho}\bar{\Psi}_{i}\gamma_{\mu}\boldsymbol{\tau}_{i}\cdot\boldsymbol{\rho}^{\mu}\Psi_{i},
\end{eqnarray}
which is basically the same as Eq.~(\ref{eq:RMFlagrangian}), except the inclusion of a vector meson $\phi_{\mu}$.
The field equations for $\sigma$, $\omega$ and $\rho$ meson are the same as that in Eqs.~(\ref{eq:eommeson1}), (\ref{eq:eommeson2}) and (\ref{eq:eommeson3}).
$\phi$ meson describes the coupling between hyperons,
and the coupling constant between $\phi$ meson and nucleons is zero, $g_{\phi N}=0$.
The Lagrangian for $\phi$ meson has exactly the same form as that of $\omega$ meson,
and hence the field equation for the two mesons are the same.

The coupling constant between nucleons and mesons are taken from Ref.~\citet{Glendenning:1991es}:
$(g_{\sigma N}/m_{\sigma})^{2}=11.79\,\textrm{fm}^{2}$,
$(g_{\omega N}/m_{\omega})^{2}=9.148\,\textrm{fm}^{2}$,
$(g_{\rho N}/m_{\rho})^{2}=9.927\,\textrm{fm}^{2}$,
$b=2.95\times 10^{-3}$, $c=-1.070\times 10^{-3}$.

The coupling constants between hyperon and mesons are fixed by the assumption of a complete SU(3) symmetry~\citep{Lopes:2013cpa}.
For $\omega$ meson, we have:
\begin{eqnarray}
		\frac{g_{\omega\Lambda}}{g_{\omega N}}&=&\frac{4+2\alpha_{v}}{5+4\alpha_{v}}\,,\\
		\frac{g_{\omega\Sigma}}{g_{\omega N}}&=&\frac{8-2\alpha_{v}}{5+4\alpha_{v}}\, , \\
	       \frac{g_{\omega\Xi}}{g_{\omega N}} &=& \frac{5-2\alpha_{v}}{5+4\alpha_{v}}\, ,
\end{eqnarray}
where $\alpha_{v}$ is a parameter.
And for the $\phi$ meson, we have:
\begin{eqnarray}
		\frac{g_{\phi N}}{g_{\omega N}} &=& \sqrt{2}\left(\frac{4\alpha_{v}-4}{5+4\alpha_{v}}\right)\,,\\
		\frac{g_{\phi\Lambda}}{g_{\omega N}}&=&\sqrt{2}\left(\frac{2\alpha_{v}-5}{5+4\alpha_{v}}\right)\,,\\
		\frac{g_{\phi\Sigma}}{g_{\omega N}}&=&\sqrt{2}\left(\frac{-2\alpha_{v}-1}{5+4\alpha_{v}}\right)\,,\\
		\frac{g_{\phi\Xi}}{g_{\omega N}}&=&\sqrt{2}\left(\frac{-2\alpha_{v}-4}{5+5\alpha_{v}}\right).
\end{eqnarray}

For $\rho$ meson,
\begin{eqnarray}
		\frac{g_{\rho\Sigma}}{g_{\rho N}}&=&2\alpha_{v}\,,\\
		\frac{g_{\rho\Xi}}{g_{\rho N}}&=&-(1-2\alpha_{v})\,,\\
		\frac{g_{\rho\Lambda}}{g_{\rho N}}&=&0\,.
\end{eqnarray}

For the $\sigma$ meson, we have:
\begin{eqnarray}
		\frac{g_{\sigma\Lambda}}{g_{\sigma N}}&=&\frac{10+6\alpha_{s}}{13+12\alpha_{s}}\,,\\
		\frac{g_{\sigma\Sigma}}{g_{\sigma N}}&=&\frac{22-6\alpha_{s}}{13+12\alpha_{s}}\,,\\
		\frac{g_{\sigma\Xi}}{g_{\sigma N}}&=&\frac{13-6\alpha_{s}}{13+12\alpha_{s}}\,, 
\end{eqnarray}
where $\alpha_{s}$ is another parameter.

We should notice that if $\alpha_{v}\neq 1$,
the coupling constant between nucleon and $\phi$ meson does not vanish,
so we need to reparameterize $g_{\omega N}$ to guarantee that the properties of nuclear matter are not affected by $\phi$ meson before the appearance of hyperon:
\begin{equation}
	g_{\omega N}\omega_{0}\rightarrow \tilde{g}_{\omega N}\omega_{0}+g_{\phi N}\phi_{0}.
\end{equation}
Using the field equations for mesons, we have:
\begin{eqnarray}
		g_{\omega N}\sum_{B}\frac{g_{\omega B}}{m_{\omega}^{2}}\rho_{B}&=&\, \tilde{g}_{\omega N}\sum_{B}\frac{\tilde{g}_{\omega B}}{m_{\omega}^{2}}\rho_{B}\nonumber\\
			&& + \, g_{\phi N}\sum_{B}\frac{g_{\phi B}}{m_{\phi}^{2}}\rho_{B}. \qquad \qquad 
\end{eqnarray}

Since this equality only holds for densities where hyperons do not appear,
the sum only runs over $N$. So we have:
\begin{equation}
	\frac{g_{\omega N}}{m_{\omega}^{2}}=\frac{\tilde{g}_{\omega N}^{2}}{m_{\omega}^{2}}+2\left(\frac{4\alpha_{v}-4}{5+4\alpha_{v}}\right)^{2}\frac{\tilde{g}_{\omega N}^{2}}{m_{\phi}^{2}}.
\end{equation}

In this work, we choose $\alpha_{v}=0.75$ and $\alpha_{s}=1.231$,
which provide an equation of state stiff enough to support a $2M_{\odot}$ neutron star at zero temperature even after the inclusion of hyperon~\citep{Lopes:2013cpa}.

The field equation and EoS of this extended RMF can be derived the same way as we have introduced in Sec.~\ref{sec:NSs},
and we can solve the eigenfrequency of the non-radial oscillation of neutron star.
The calculated oscillation eigenfrequencies of different modes of neutron star are shown in Table.~\ref{tab:gmode_RMF_II}.
The eigenfrequency is almost the same as the ones shown in Table.~\ref{tab:gmode} and ~\ref{tab:fpmode}.
The main difference of the RMF described in Sec.~\ref{sec:NSs} and the one introduced here is the way of including hyperon.
However, since the mass of the proto-neutron star is relatively small,
and the central density is not high enough for the hyperons to appear,
the EoSs of different hadron model are very close to each other,
because all these models are calibrated by experimental data at low densities.

\begin{table}[!t]
\begin{center}
	\caption{The eigenfrequency of oscillation for NS constructed with extended RMF model.}
\label{tab:gmode_RMF_II}
\begin{tabular}{c|ccc}
\hline \hline                    %\vspace{0.1cm}
 & \multicolumn{3}{c}{Neutron Star}    \\\hline
Mode    & $t\!=\!100$&$t\!=\!200$&$t\!=\!300$\\
	$_2g_2$ & 432.7& 449.8 & 444.6    \\
	$_2g_1$ & 718.8& 762.6 & 765.7   \\
	$_2f$   & 1116 & 1132  & 1177    \\
	$_2p_1$ & 2429 & 2590  & 2697    \\
\hline \hline
\end{tabular}
\end{center}
\end{table}

\subsection{NJL model with vector interactions}\label{sec:stiffNJL}
For the stiff quark model, we adopt here an NJL model with vector interaction~\citep{Chu:2014pja,Chu:2016ixv,Chu:2017bmr}.

\begin{equation}
	\mathcal{L}=\bar{\psi}\left(i\gamma_{\mu}\partial^{\mu}-\hat{m}_{0}\right)+\mathcal{L}_{S}+\mathcal{L}_{V}+\mathcal{L}_{\textrm{Hooft}},
\end{equation}
where $\mathcal{L}_{S}$ is the scalar interaction, the $\mathcal{L}_{V}$ is the vector interaction,
and $\mathcal{L}_{\rm Hooft}$ is the 't Hooft term.
$\psi$ is the quark fields and $\psi=(\psi_{u},\psi_{d},\psi_{s})^{T}$.
$\hat{m}_{0}$ is the matrix of the current quark masses $\hat{m}_{0}=\textrm{diag}(m_{u},m_{d},m_{s})$.

For the scalar interaction, we have:
\begin{equation}
	\mathcal{L}_{S}=G_{S}\sum_{a=0}^{8}\left[\left(\bar{\psi}\tau_{a}\psi\right)^{2}+\left(\bar{\psi}i\gamma_{5}\tau_{a}\psi\right)^{2}\right]\,,
\end{equation}
and for the 't Hooft term, we have:
\begin{eqnarray}
\mathcal{L}_{\textrm{Hooft}} =-K\left[\textrm{det}_{f}\left(\bar{\psi}(1+\gamma_{5})\psi\right)\right. \nonumber\\
+\left.\textrm{det}_{f}\left(\bar{\psi}(1-\gamma_{5})\psi\right)\right],
\end{eqnarray}
which is the same as in Sec.~\ref{sec:NJL}.

For the vector interaction, in order to construct a pure quark star over $2M_{\odot}$,
we use the interaction introduced in \citet{Chu:2014pja,Chu:2016ixv,Chu:2017bmr}:
\begin{equation}
	\mathcal{L}_{V}=-g_{_V}\left(\bar{\psi}\gamma^{\mu}\psi\right)^{2}
	-G_{IV}\left[(\bar{\psi}\gamma^{\nu}\boldsymbol{\tau}\psi)^{2}+\left(\bar{\psi}\gamma^{\nu}\gamma_{5}\boldsymbol{\tau}\psi\right)^{2}\right],
\end{equation}
where $g_{_V}$ is the vector coupling constant, and $G_{IV}$ is the vector-isovector coupling constant.

In the mean-field approximation, the thermodynamic potential becomes:
\begin{eqnarray}
\Omega_{QM}&=&-2N_{c}\int_{0}^{\Lambda}\frac{\textrm{d}^3 p}{(2\pi)^{3}}\bigg\{\beta^{-1}\ln\Big[1+e^{-\beta(E_{i}^{}(k) - \tilde{\mu}_{i}^{})}\Big]\nonumber\\
&&+\beta^{-1}\ln\Big[1+e^{-\beta(E_{i}^{}(k) + \tilde{\mu}_{i}^{})}\Big]\,+E_{i}^{} \bigg\}\nonumber\\
&&+2G_{S}\left({\phi_{u}}^{2} +{\phi_{d}}^{2}+{\phi_{s}}^{2}\right) - 4K \phi_{u}^{}\,\phi_{d}^{}\,\phi_{s}^{}\nonumber\\
&&-\left[g_{_V}(\rho_{u}+\rho_{d}+\rho_{s})^2+G_{IV}(\rho_{u}-\rho_{d})^2\right]\nonumber\\
&&+C\,,\label{eq:CHUpotential}
\end{eqnarray}
which is basically the same as in Eq.~(\ref{eq:NJLpoten}) except that the fourth line is the contribution from vector interaction.
$\rho_{i}$ is the number density of the quark flavor $i$.
The constant $C$ in the last line is fixed by requiring $\Omega_{QM}=0$ in the vacuum.

The $\tilde{\mu}_{i}$ in Eq.~(\ref{eq:CHUpotential}) is the effective chemical potential, and is given by:
\begin{equation}
\tilde{\mu}_{i}=\mu_{i}-2g_{_V}\sum_{i=u,d,s}\rho_{i}-2G_{IV}\tau_{3i}(\rho_{u}-\rho_{d}),
\end{equation}
where $\mu_{i}$ is the chemical potential for quarks, and $\tau_{3i}$ is the isospin of quarks:
$\tau_{3u}=1$, $\tau_{3d}=-1$ and $\tau_{3s}=0$.

The dispersion relation, the constituent quark mass, the quark condensate and the quark number density
is the same as in Eqs.~(\ref{eq:NJLdispersion}), (\ref{eq:constituentmass}), (\ref{eq:condensate}) and (\ref{eq:NJLdensity}), respectively,
except that all the quark chemical potential should be replaced by the corresponding effective chemical potential.

The pressure is the minus of the thermodynamic potential $p_{QM}=-\Omega_{QM}$.
The entropy density $S_{QM}$ is the partial derivative of pressure with respect to temperature,
    which has the same form as in Eq.~(\ref{eq:NJLentropy}) except that the quark chemical potential should be replaced by the corresponding effective chemical potential.
The energy density of the quark matter can be obtained from the thermodynamic relation:
\begin{equation}
\varepsilon_{QM}=TS_{QM}+\sum_{i=u,d,s}\mu_{i}\rho_{i}-p_{QM}.
\end{equation}

The parameter we use here is~\citep{Chu:2014pja}: $\Lambda = 631.4\,\textrm{MeV}$,
$G_{S}\Lambda^{2}=1.835$,
$K\Lambda^{5}=9.29$,
$g_{_V}=1.1G_{S}$,
$G_{IV}=0$,
$m_{u}=m_{d}=5.5\,\textrm{MeV}$,
$m_{s}=135.7\,\textrm{MeV}$,
$m_{e}=0.511\,\textrm{MeV}$,
and $m_{\mu}=105.7\,\textrm{MeV}$.

\begin{figure}[t]
\begin{center}
\includegraphics[width=0.45\textwidth]{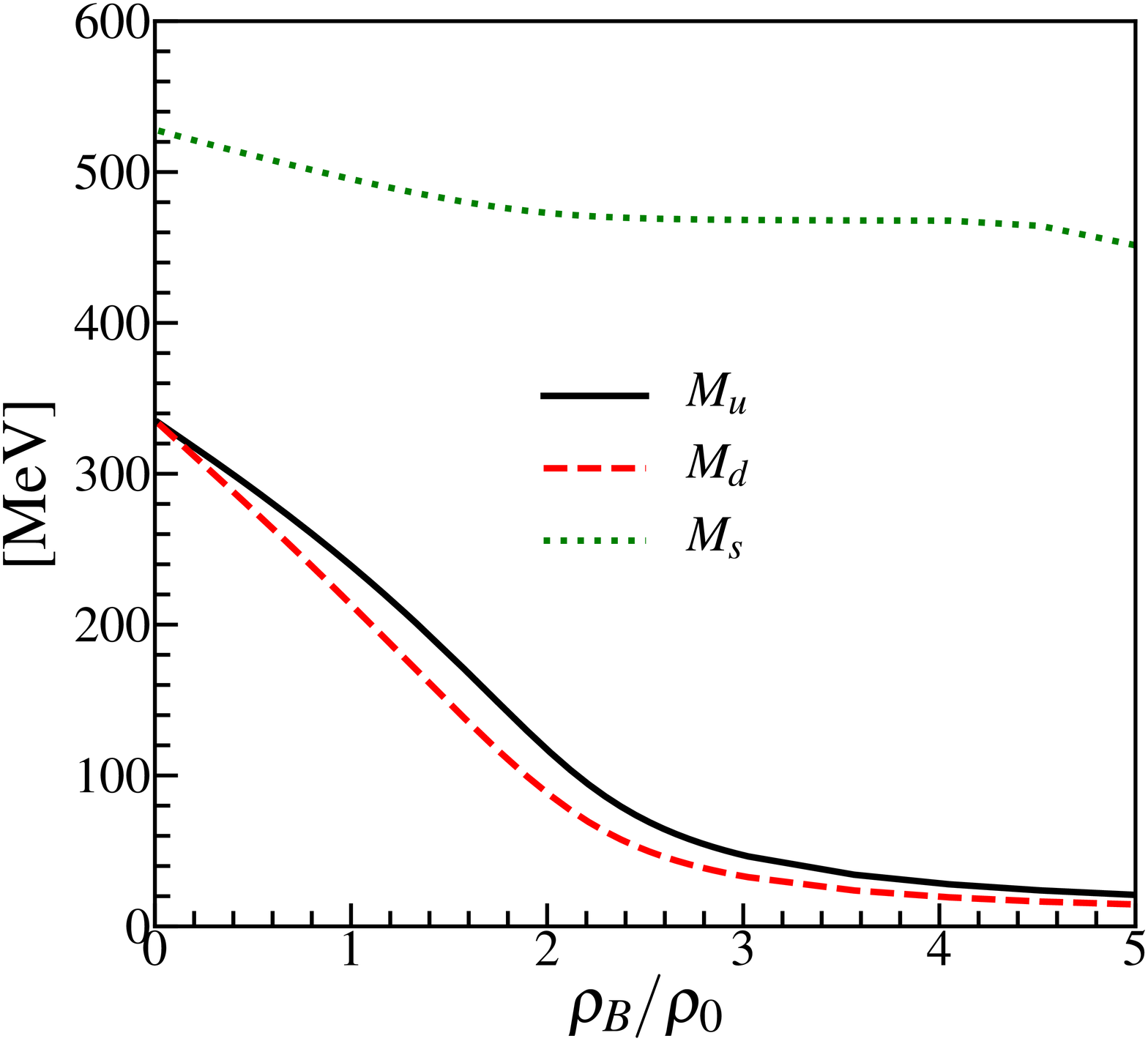}
\vspace*{-3mm}
	\caption{Same as Fig.~\ref{fig:quarkmass}, but using NJL model with vector interaction.}
	\label{fig:quarkmass2}
\end{center}
\end{figure}

The calculated constituent quark mass at zero temperature is shown in Fig.~\ref{fig:quarkmass2}.
Comparing Fig.~\ref{fig:quarkmass} and Fig.~\ref{fig:quarkmass2},
we see that the constituent quark mass at certain baryon density is almost the same for the two different NJL model.
However, since the EoS for NJL model with vector interaction is stiff,
for a given mass of the SQS, the center density is smaller,
and the constituent quark mass in the SQS will be generally larger.

\begin{figure*}[htb]
\begin{center}
\includegraphics[width=0.88\textwidth]{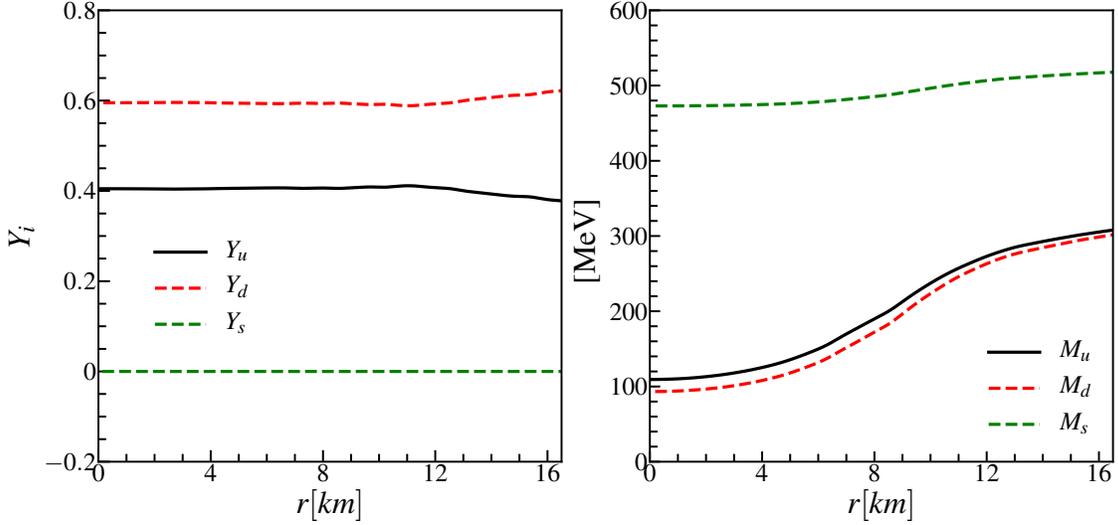}
\vspace*{-3mm}
	\caption{The same as Fig.~\ref{fig:Yimass}, except that this is for NJL model with vector interaction.}
\label{fig:R_Mi_Yi2}
\end{center}
\end{figure*}

The calculated quark fraction and current quark mass as a function of SQS radius is shown in Fig.~\ref{fig:R_Mi_Yi2}.
The quark matter is mainly consisted of $u$ and $d$ quark,
and the $s$ quark fraction is almost zero throughout the whole star.
However, unlike shown in Fig.~\ref{fig:Yimass},
the constituent quark mass of $u$ and $d$ quark ranges from $260\,\textrm{MeV}$ in the center to about $310\,\textrm{MeV}$ on the surface,
which is non-neglectable and is not highly relativistic.
This means the eigenfrequency of the non-radial oscillation should be larger compared to that of NJL model without vector interaction.

The calculated result of oscillation frequencies using NJL model with vector interaction is shown in Table.~\ref{tab:NJLgmode2}.
Comparing with Table.~\ref{tab:NJLgmode}, we see that the eigenfrequency is significantly larger when the vector interaction is included.
This is not surprising since, as we have argued, the constituent mass of the component quarks is larger with vector interaction than that without.
However, the constituent quark mass is still much smaller than the mass of hadron,
and consequently, the g-mode oscillation frequency for SQS is still much smaller than that of NS.

\begin{table}[htb]
\begin{center}
	\caption{The same as Table.~\ref{tab:NJLgmode}, except that we use the NJL model with vector interaction.}
\label{tab:NJLgmode2}
\begin{tabular}{c|ccc}
\hline \hline                    %\vspace{0.1cm}
$g$-mode & \multicolumn{3}{c}{Strange Quark Star}    \\\hline
Radial order  & $t\!=\!100$&$t\!=\!200$&$t\!=\!300$ \\
 $_{2}g_2$ & 205.0  & 208.7 & 190.7   \\
 $_{2}g_1$ & 349.8  & 372.1 & 362.6   \\
  $_{2}f$  & 1163.3 & 1207.1& 1258.8  \\
  $_{2}p_1$& 4218.8 & 4134.3& 4121.9  \\
%  $_{2}p_2$& 6552.7 & 6433.3& 6438.3  \\
\hline \hline
\end{tabular}
\end{center}
\end{table}

On the other hand, the eigenfrequencies of the $f$- and $p$-mode oscillations are of the same order as that of NSs.
Unlike $g$-mode oscillation, the $f$-mode and $p$-mode oscillations originate from the pressure inside the star,
and their eigenfrequencies are closely related to the compactness of the star~\citep{Chirenti:2015dda}.

\begin{figure}[!htbp]
\begin{center}
\includegraphics[width=0.45\textwidth]{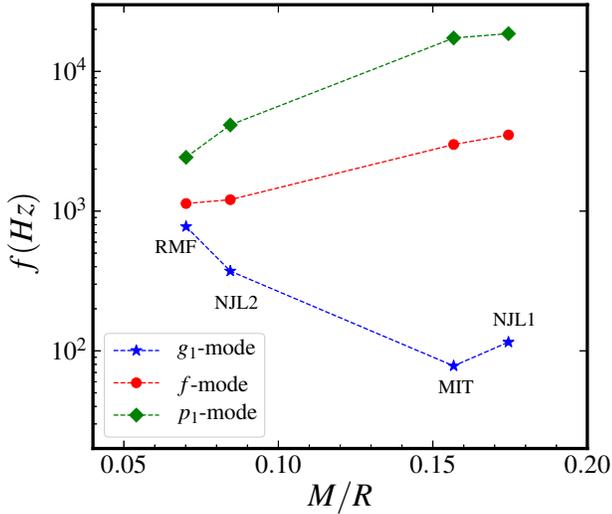}
\vspace*{-2mm}
\caption{
	The relation between compactness $M/R$ and the eigenfrequency of different modes for different models.
	$t=200$ms is chosen as a representative of time.
	The eigenfrequencies with same compactness correspond to the same model.
	The RMF model is described in Sec.~\ref{sec:NSs},
	the MIT model is described in Sec.~\ref{sec:MIT},
	the NJL1 model is described in Sec.~\ref{sec:NJL},
	and the NJL2 model is described in Sec.~\ref{sec:stiffNJL}.
}\label{fig:f_MR}
\end{center}
\end{figure}

In Fig.~\ref{fig:f_MR}, we show the relation between eigenfrequencies and compactness for different models.
We choose $t=200\;$ms as a representative of time,
and the mass for different models are all $0.95M_\odot$ according to our assumption.
However, since different models lead to different radius,
the compactness for different models are different.
Since the RMF model described in Sec.~\ref{sec:NSs} and ~\ref{sec:stiffRMF} give almost the same result,
we only present the result using RMF model in Sec.~\ref{sec:NSs} in this figure.
The MIT model is described in Sec.~\ref{sec:MIT},
and NJL1 and NJL2 model is described in Sec.~\ref{sec:NJL} and \ref{sec:stiffNJL}, respectively.

It is apparent from the figure that MIT model and NJL1 model have larger compactness,
i.e. their radius is smaller, and the resulting $f$- and $p$- mode eigenfrequencies is larger.
The SQS from NJL2 model, however, has a larger radius, so its compactness is closer to the NSs,
and the $f$- and $p$-mode eigenfrequency are also closer to the NSs.

\section{Damping through gravitational wave}\label{sec:damping}

In previous sections, we have shown that the $g$-mode oscillation frequencies are significantly different for NSs and SQSs.
However, in order to observe this difference on earth by GW detection,
we also require that the energy emission through GW is efficient.

%Apart from the energy of the oscillation, the damping of the mode is also essential.
%According to \citet{Reisenegger:1992APJ}, there are three types of damping mechanism:
%they are the {\it relaxation toward chemical equilibrium},
%the {\it viscous} damping and the {\it gravitational wave} damping.
%Among these mechanisms, the damping through gravitational wave is of most interest in this work.

The timescale $\tau$ associated with any dissipative process is:
\begin{equation}
	\tau=\frac{2E}{P},
\end{equation}
where $E$ is the energy of the oscillation,
and $P=\textrm{d}E/\textrm{d}t$ is the power of the dissipation.

For a star oscillates at an eigenfrequency $\omega$, the mode energy is given by Eq.~(\ref{eqn:ModeEnergy}),
and the radiated power (in the weak-gravity approximation) is~\citep{Reisenegger:1992APJ}:
\begin{equation}
	P=\frac{G}{8\pi c^{2l+1}}\frac{(l+1)(l+2)}{(l-1)l}\left[\frac{4\pi\omega^{l+1}}{(2l+1)!!}\int_0^{R}\rho_{1}r^{l+2}\textrm{d}r\right]^2,
\end{equation}
where $\rho_{1}$ is defined by:
\begin{equation}\label{eq:deltarho}
	\delta\rho(r,\theta,\phi,t)=\rho_{1}Y_{l}^{m}(\theta,\phi)\textrm{e}^{-i\omega t}.
\end{equation}

According to Eqs.~(\ref{eq:drho}) and (\ref{eq:displa}), $\rho_1$ can be expressed as:
\begin{equation}\label{eqn:rho1rho0}
\frac{\rho_1}{\rho}=\frac{\omega_{BV}^{2}}{g}\eta_{r}+\frac{\omega^2 r}{c_{s}^{2}}\eta_{\bot}.
\end{equation}

Therefore, we have:
\begin{eqnarray}
\tau&=&\frac{c^{2l+1}}{G}\frac{1}{2\pi\omega^{2l}}\frac{[(2l+1)!!]^2\times (l-1)l}{(l+1)(l+2)}\nonumber\\
& & \times \frac{\int_0^R \td r\rho r^2\left[\eta_r^2+l(l+1)\eta_{\bot}^2\right]}
{\left\{\int_0^R\td r\rho r^{l+2}\left[\frac{\omega_{BV}^2}{g}\eta_{r}+\frac{\omega^2 r}{c_{s}^{2}}\eta_{\bot}\right]\right\}^{2}} \, .
\end{eqnarray}

The calculated damping times for different models and different modes are shown in Table~\ref{tab:tau}.
As can be seen from the table,
for NSs, the damping time of $l=2$, $n=1$ $g$-mode oscillation is of the order of a hundred seconds.
For soft SQSs, i.e., SQSs with MIT or NJL1 model,
the damping time of $_{2}g_{1}$ mode is from several tens to several thousands of years.
This means that the GW from the $g$-mode oscillation is extremely unlikely to be detected on earth,
since most of the oscillation energy will be dissipated from other processes such as viscous or relaxation towards chemical equilibrium.
For SQSs described with NJL2 model,
the $_{2}g_{1}$ mode damping time is about one order larger than the NSs,
but is significantly smaller than that of soft SQSs.
This means that the $g$-mode oscillation of stiff SQSs is not impossible to be detected on earth,
although it will be much more difficult than the NSs.

{%\color{red}
In order to make a rough estimation of the possibility of detecting the GW from NJL2 model,
we adopt the commonly used quadrupole approximation to calculate GW strain~\citep{Thorne:1980ru,Maggiore:GWv1}.

The metric perturbation in transverse-traceless (TT) gauge is~\citep{Finn:1990qf,Ott:2003qg,Szczepanczyk:2021bka}.
\begin{equation}
h^{\textrm{\tiny TT}}_{i,j}(t,\vec{x})=\frac{1}{D}\ddot{Q}_{ij}(t-D/c,\vec{x}),
\end{equation}
where $i$ and $j$ are the spatial index of Cartesian coordinate, $i,j=1,2,3$.
$D$ is the distance from source to the observer, $c$ is speed of light,
and the dots refer to time derivative.
$Q_{ij}$ is the traceless quadrupole moment:
\begin{equation}\label{eq:TracelessQuadrupoleMoment}
Q_{ij}=\frac{2G}{c^4}\hat{\Lambda}(\vec{n})\int\td^3x\rho(t,\vec{x})(x_ix_j-\frac{1}{3}\delta_{ij}|\vec{x}|^2),
\end{equation}
where $G$ is the gravity constant,
$\rho$ is the matter density,
$\vec{n}$ is the unit vector pointing from source to observer,
and $\hat{\Lambda}$ is the TT projection operator.
For a rough estimation, we will take $\hat{\Lambda}=1$,
meaning that the source and the detector are in perfect direction that the GW can be maximally observed.

The metric perturbation can be decomposed according to its polarization:
\begin{equation}
h^{\textrm{\tiny TT}}_{ij} = h_{+}\vec{e}_{+}+h_{\times}\vec{e}_{\times},
\end{equation}
where $\vec{e}_+$ and $\vec{e}_\times$ are the unit vector of plus and cross polarization.

For axisymmetric supernova explosion, the cross polarization is zero, $h_\times=0$.
The $Q_{ij}$ is diagonal and traceless, $Q_{11}=Q_{22}=-\frac{1}{2}Q_{33}$.
The GW strain $h_+$ is then~\citep{Finn:1990qf}:
\begin{equation}
h_+= \frac{3\sin^2\theta}{2D} \ddot{Q}_{33},
\end{equation}
where $\theta$ is the inclination angle.
Again, for a rough estimation, we will take $\sin\theta=1$ for maximally observing the GW.

The mass distribution is given by:
\begin{equation}
\rho(t,\vec{x})=\rho_0(r)+\delta\rho(r,\theta,\phi,t),
\end{equation}
where $\rho_0$ is the static configuration and $\delta\rho$ is given by Eq.~(\ref{eq:deltarho}).
Substituting into Eq.~(\ref{eq:TracelessQuadrupoleMoment}), we have:
\begin{eqnarray}
\ddot{Q}_{33}&=&\frac{2G}{c^4}\int r^2\td r \sin\theta\td\theta \td\phi \frac{\td^2\delta\rho}{\td t^2}(r^2\sin^2\theta-\frac{1}{3}r^2)\nonumber\\
%&=&-\omega^2\frac{2G}{c^4}\int r^2\td r \sin\theta\td\theta\td\phi\delta\rho(t,r,\theta,\phi)(r^2\sin^2\theta-\frac{1}{3}r^2)\nonumber\\
%&=&-\omega^2\frac{2G}{c^4}\int r^2\td r \sin\theta\td\theta\td\phi\rho_1(r)Y_{lm}(\theta,\phi)(r^2\sin^2\theta-\frac{1}{3}r^2)\e^{-i\omega t}\nonumber\\
&=&|\ddot{Q}_{33}|\e^{-i\omega t},
\end{eqnarray}
where
\begin{equation}
|\ddot{Q}_{33}| =\omega^2\frac{2G}{c^4}\frac{4\sqrt{\pi/5}}{3}\int r^4\td r \rho_1(r),
\end{equation}
where we have used Eq.~(\ref{eq:deltarho}), and taken $l=2$, $m=0$.
The amplitude of the strain can then be defined as $|h_+|=3|\ddot{Q}_{33}|/2D$.

We notice that the value of $\rho_1$ cannot be fixed,
since the amplitude of the oscillation is determined by the detail of the supernova explosion
and is beyond the scope of this paper.
Therefore, the metric perturbation $h_{+}$ cannot be fixed.
However, we can compare the GW strain for NS and SQS with proper normalization condition.
We require that the amplitude of the $n=1$ $g$-mode oscillation is the same for NS and SQS at $t=200$ms,
and we calculate the ratio between the strains.

Our calculation shows that $|h_+^{\textrm{\tiny RMF1}}|/|h_+^{\textrm{\tiny NJL2}}|=5.32$,
if $D$ is the same for NS and SQS.
This means that for a rough estimation,
the GW for PSQS is about 5 times weaker than the PNS,
if PNS and PSQS  are at the same distance and oscillate at the same amplitude,

There have been attempts for the search of supernova GW~\citep{LIGOScientific:2016jvu,LIGOScientific:2019ryq}.
Although the corresponding signal has not been detected so far,
it is possible to be detected in current generation of detectors~\citep{Szczepanczyk:2021bka,Gossan:2015xda}.
However, according to our estimation,
the GW from PSQS may be delayed to future generation of detector,
since the GW strain is much weaker.
}

As for the $f$- and $p$- mode, the damping time is very small,
and the GW energy emission is efficient.
For NSs and soft SQSs, the $f$-mode frequencies are very different and can be distinguished.
However, for stiff SQSs, the $f$-mode frequency is very close to the NSs,
and the component cannot be determined through GW detection.

Also, the frequencies of $f$- and $p$-mode oscillation is not in the range that
 current ground based gravitational wave detector is sensitive.
{%\color{red}
It has also been suggested that the g-mode oscillation is most likely to be excited in supernova explosion
~\citep{Burrows:2005dv,Ott:2006qp}
}

Therefore, among all these modes, the $g$-mode is still the most possible one to provide insights on the component particle of the compact object.

\begin{table*}[htb]
\begin{center}
\caption{
	The damping timescale through gravitational wave emission for NSs and SQSs with different models.
$t=200\;$ms is chosen as a representative of time.
RMF1 is described in Sec.~\ref{sec:NSs},
RMF2 is described in Sec.~\ref{sec:stiffRMF},
MIT is described in Sec.~\ref{sec:MIT},
NJL1 is described in Sec.~\ref{sec:NJL},
and NJL2 is described in Sec.~\ref{sec:stiffNJL}.
}
\label{tab:tau}
\begin{tabular}{c|cc|ccc}
\hline \hline                    %\vspace{0.1cm}
      &\multicolumn{2}{c|}{NS} & \multicolumn{3}{c}{SQS}  \\
\hline
Modes        & RMF1                 &RMF2                  &MIT                   & NJL1                 &   NJL2                \\
\hline
$_{2}g_{2}$  & $4.12\times 10^{3}$  & $4.31\times 10^{3}$  & $7.81\times 10^{12}$ & $6.19\times 10^{11}$ & $2.03\times 10^{6}$   \\
$_{2}g_{1}$  & $1.51\times 10^{2}$  & $1.14\times 10^{2}$  & $2.01\times 10^{11}$ & $1.67\times 10^{9}$  & $7.84\times 10^{3}$   \\
$_{2}f$      &     $2.92$           &  $3.08$              & $3.13\times 10^{1}$  & $8.39$               & $3.36\times 10^{-1}$  \\
$_{2}p_{1}$  & $1.65\times 10^{-1}$ & $1.54\times 10^{-1}$ & $2.14\times 10^{-4}$ & $1.85\times 10^{-4}$ & $2.31\times 10^{-2}$  \\
\hline \hline
\end{tabular}
\end{center}
\end{table*}

\section{\label{sec:sum}Conclusions}

In this work we have studied the nonradial oscillations of newly born NSs and SQSs.
The relativistic nuclear field theory with hyperon degrees of freedom is employed to describe the equation of state for the stellar matter in NSs,
while both the MIT bag model and the Nambu--Jona-Lasinio model are adopted to construct the configurations of the SQSs.
We find that there are no hyperons in the newly born NSs, and it is also found that the $g$-mode eigenfrequencies of newly born SQSs are much lower than those of NSs,
no matter which model we choose to describe the newly born SQSs, which implies that the conclusion is model independent.

Note, however, that there are some differences between the strange quark matter described by the bag model and that by the NJL model.
In the MIT bag model, all the three flavor quarks are relativistic, and the maximal particle mass is the current mass of strange quarks,
about $150\,\mathrm{MeV}$.
In the NJL model without vector interaction, quark masses are generated (or dressed) through the DCSB, and the $u$ and $d$ quarks are still relativistic,
but the strange quark has large constituent mass.
Because of the large constituent mass of the $s$ quark, its abundance in the charge-neutral SQSs is very low, therefore,
all the constituents of SQSs are still relativistic, which results in low $g$-mode eigenfrequencies of newly born SQSs.

We have also calculated the $g$-mode oscillation frequency using models that is stiff enough to support $2M_{\odot}$ compact object.
For stiff hadron model, the $g$-mode oscillation frequency is almost the same as that of the soft one,
    since the main difference of the two models is the way to introduce hyperons,
    and the density inside the proto-neutron star is not large enough to generate hyperon.
For stiff quark model, however, the $g$-mode oscillation frequency is significantly larger than the soft quark model.
The reason is that since we have fixed the mass of the proto strange quark star,
    a stiffer EoS corresponds to a smaller inner density,
    and the constituent quark mass is larger at smaller densities, which is a general feature of QCD.
As we have stated, the eigenfrequency of the $g$-mode oscillation is closely related to the mass of the component particle,
   and a larger component particle mass will lead to a larger oscillation frequency.
However, even for this extreme quark model, the $g$-mode oscillation frequency for SQS is still significantly smaller than that of NS.
Therefore, the $g$-mode oscillation frequency can still serve as a great criterion to identify SQS from NS.

We have also investigated other modes of nonradial oscillations of newly born NSs and SQSs, such as the $f$- and $p$-modes.
We find that eigenfrequencies of the $f$- and $p$-modes are larger than those of $g$-mode.
For SQSs with soft EoS, the $f$- and $p$-modes eigenfrequencies are larger than those of the newly born NSs.
However, for SQSs with stiff EoS, $f$- and $p$-mode eigenfrequencies are close to the NSs,
because the compactness of the SQS with stiff EoS is close to the NSs.

In order to further study the possibility of distinguishing the component of the newly born object,
we study the damping time through GW emission for different modes.
For NSs, the damping time of $g$-mode oscillation is not very large,
which means that there should be a considerable amount of energy that is transferred from oscillation to the GW emission,
and it is possible for the detectors on earth to observe it.
For SQSs with soft EoS, however, the damping times of the $g$-mode oscillation are extremely large,
and there will be very little energy emitted through GW.
However, for SQSs with stiff EoS, although the $g$-mode damping time is larger than NSs,
it is significantly smaller than that of SQSs with soft EoS.
{%\color{red}
A rough estimation indicate that the GW strain is about 5 times weaker for stiff SQS than NS,
}
and it is still possible to be detected on earth for future generation of detector.

In the light of the first direct observation of gravitational waves~\citep{LIGOScientific:2016aoc},
and the $g$-mode oscillations of the type II supernovae core serving as potential, efficient sources of gravitational waves~\citep{Ott:2006qp},
it is promising to employ the gravitational waves to identify the QCD phase transition in compact stars, {\it i.e.}, high density strong interaction matter.

\begin{acknowledgments}

The work was supported by the National Natural Science Foundation of China under Contracts No.~11775041, 
and the Fundamental Research Funds for the Central Universities under Contract No. DUT16RC(3)093.
\end{acknowledgments}

%\bibliography{Ref-PNS_Oscillation}{}
%\bibliographystyle{aasjournal}

\end{document}